\documentclass[a4paper,11pt]{article}
\pdfoutput=1 
\usepackage{jheppub} 
\usepackage[T1]{fontenc} 

\usepackage{epsfig}
\usepackage{graphicx}
\usepackage{dcolumn}
\usepackage{bm}
\usepackage{ltablex,booktabs}
\usepackage{overpic}
\usepackage{subfigure}
\usepackage{float}
\usepackage{color}
\usepackage{amsmath}
\usepackage{mathcomp}
\usepackage{mathrsfs}
\usepackage{multirow}
\usepackage{rotating}
\usepackage{amssymb}
\usepackage{gensymb}
\usepackage{amsmath}
\usepackage{tabularx}
\usepackage{epstopdf}

\title{Amplitude analysis and branching-fraction measurement of \boldmath $D_{s}^{+} \to K^0_{S}\pi^{+}\pi^{0}$}
\collaborationImg{\includegraphics[width=0.15\textwidth, angle=90]{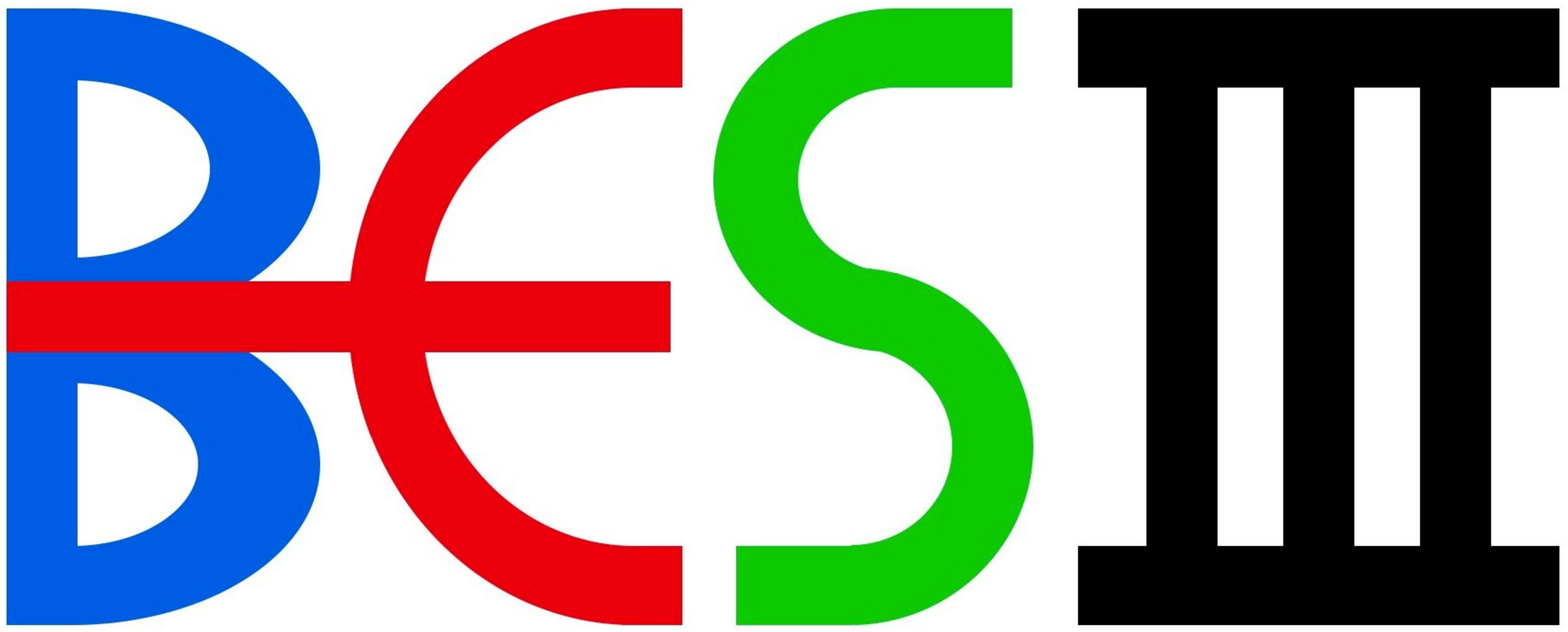}}
\collaboration{BESIII Collaboration}

\author{
M.~Ablikim$^{1}$, M.~N.~Achasov$^{10,b}$, P.~Adlarson$^{67}$, S. ~Ahmed$^{15}$, M.~Albrecht$^{4}$, R.~Aliberti$^{28}$, A.~Amoroso$^{66A,66C}$, M.~R.~An$^{32}$, Q.~An$^{63,49}$, X.~H.~Bai$^{57}$
% ...
\\
\vspace{0.2cm} {\it
$^{1}$ Institute of High Energy Physics, Beijing 100049, People's Republic of China\\
% ...
$^{4}$ Bochum Ruhr-University, D-44780 Bochum, Germany\\
% ...
$^{10}$ G.I. Budker Institute of Nuclear Physics SB RAS (BINP), Novosibirsk 630090, Russia\\
% ...
$^{15}$ Helmholtz Institute Mainz, Staudinger Weg 18, D-55099 Mainz, Germany\\
% ...
$^{28}$ Johannes Gutenberg University of Mainz, Johann-Joachim-Becher-Weg 45, D-55099 Mainz, Germany\\
% ...
$^{32}$ Liaoning Normal University, Dalian 116029, People's Republic of China\\
% ...
$^{49}$ State Key Laboratory of Particle Detection and Electronics, Beijing 100049, Hefei 230026, People's Republic of China\\
% ...
$^{57}$ University of Jinan, Jinan 250022, People's Republic of China\\
% ...
$^{63}$ University of Science and Technology of China, Hefei 230026, People's Republic of China\\
% ...
$^{66}$ University of Turin and INFN, (A)University of Turin, I-10125, Turin, Italy; (B)University of Eastern Piedmont, I-15121, Alessandria, Italy; (C)INFN, I-10125, Turin, Italy\\
$^{67}$ Uppsala University, Box 516, SE-75120 Uppsala, Sweden\\
% ...
$^{b}$ Also at the Novosibirsk State University, Novosibirsk, 630090, Russia\\
% ...
}
}
\date{\today}
\abstract{
Utilizing a data set corresponding to an integrated luminosity of 6.32~$\rm fb^{-1}$,
recorded by the BESIII detector at  center-of-mass energies between 4.178 and 4.226~GeV, we perform an amplitude analysis of the decay
$D_{s}^{+} \to K_{S}^{0}\pi^{+}\pi^{0}$ and determine the relative fractions and
phase differences of different intermediate processes, which include $K_{S}^{0}\rho(770)^{+}$,
$K_{S}^{0}\rho(1450)^{+}$, $K^{*}(892)^{0}\pi^{+}$, $K^{*}(892)^{+}\pi^{0}$, and $K^{*}(1410)^{0}\pi^{+}$.
Using a double-tag technique, and making an efficiency correction that relies on our knowledge
of the phase-space distribution of the decays coming from the amplitude analysis,
 the absolute branching
fraction is measured to be
$\mathcal{B}(D_{s}^{+} \to K_{S}^{0}\pi^{+}\pi^{0})=(5.43\pm0.30_{\text{stat}}\pm 0.15_{\text{syst}})\times 10^{-3}$.
}

\arxivnumber{}
\begin{document}
\maketitle
\flushbottom


\section{Introduction}
Knowledge of  $D_{s}^{\pm}$  decay properties   are vital input for studies of
the $B^0_s$  hadron, whose decay channels are dominated by the final states 
involving $D_{s}^{\pm}$ mesons~\cite{PDG}. Furthermore, hadronic $D_{s}^{\pm}$ 
decays probe the interplay of short-distance weak-decay matrix elements and 
long-distance QCD interactions, and the measured branching fractions (BFs) 
provide valuable information concerning the amplitudes and phases that the 
strong force induces in the decay process~\cite{PRD79-034016, PRD81-074021, PRD84-074019}. 
The singly Cabibbo-suppressed (SCS) decay $D_{s}^{+} \to K^{0}\pi^{+}\pi^{0}$ 
has a large BF of the order of $10^{-2}$~\cite{PDG}. This decay, therefore, is 
often used as a reference channel for the other decays of $D_{s}^{\pm}$ mesons.
Accurate knowledge of its substructure is essential to reduce the systematic 
uncertainties in those analyses using this channel. To date, there have been 
few measurements of charge-parity asymmetries $A_{\rm CP}$ in SCS $D_{s}^{\pm}$ 
decay modes in general and none for the mode discussed here.

An amplitude analysis of the $D^+_s$ decay to a three-body pseudoscalar meson 
final state is a powerful tool for studying the vector-pseudoscalar channels of
the SCS $D^+_s$ decay. Table~\ref{tab:BF_prediction} shows the current measured 
values and theoretical predictions, in various models, for the BFs of
$D^+_s\to K_{S}^{0}\rho^+$, $K^{*}(892)^{0}\pi^+$, and $K^{*}(892)^{+}\pi^0$
($\rho^+$ denotes $\rho(770)^+$ throughout this paper).
References~\cite{YLWu} and \cite{HYCheng} took into account quark flavor SU(3)
symmetry and its breaking effects. Reference~\cite{PRD84-074019} used a 
generalized factorization method considering the resonance effects in the pole 
model for the annihilation contributions and introducing large strong phases 
between different topological diagrams. More precise  experimental results are 
required to validate or falsify  these theoretical predictions.

The CLEO collaboration has reported a measurement of
$\mathcal{B}(D_{s}^{+} \to K^{0}\pi^{+}\pi^{0}) = (1.00\pm 0.18)\%$~\cite{CLEO-BF},
using 600~pb$^{-1}$ of $e^+e^-$ collisions recorded at a 
center-of-mass energy~($\sqrt{s}$) of 4.17~GeV. In this paper, by using
6.32~$\rm fb^{-1}$ of  data  collected with the BESIII detector  at 
$\sqrt{s}=4.178$-$4.226$~GeV, we perform the first amplitude analysis of
$D^+_s\to K^0_S\pi^+\pi^0$ and improve the measurement of its absolute BF.
\begin{table*}[htp]
 \renewcommand\arraystretch{1.25}
\begin{center}
  \caption{Summary of $D_s^+$ decays to a vector and pseudoscalar meson, 
    showing the measured BFs and theoretical predictions from various models ($\times 10^{-3}$).}
\begin{tabular}{lcccc}
\hline
Channel                               & PDG~\cite{PDG}     & Y.L.~Wu {\it et~al.}~\cite{YLWu} & H.Y.~Cheng {\it et~al.}~\cite{HYCheng}& F.S.~Yu {\it et~al.}~\cite{PRD84-074019}\\
\hline
$K^{0}\rho^{+}$         & --                 & $9.1 \pm 7.7$                     & $11.47 \pm 0.48$                       & $7.5 \pm 2.1$                \\
$K^{*}(892)^{0}\pi^{+}$ & $2.13 \pm 0.36$    & $3.3 \pm 3.5$                     & $\phantom{0}3.65 \pm 0.24$             & $1.5 \pm 0.7$                \\
$K^{*}(892)^{+}\pi^{0}$ & --                 & $1.3 \pm 1.3$                     & $\phantom{0}1.02 \pm 0.07$             & $0.1 \pm 0.1$                \\
\hline
\end{tabular}
\label{tab:BF_prediction}
\end{center}
\end{table*}

\section{Detector and data sets}
\label{sec:detector_dataset}
The BESIII detector is a magnetic spectrometer~\cite{Ablikim:2009aa, Ablikim:2019hff}
located at the Beijing Electron Positron Collider~(BEPCII)~\cite{Yu:IPAC2016-TUYA01}.
The cylindrical core of the BESIII detector consists of a helium-based
multilayer drift chamber~(MDC), a plastic scintillator time-of-flight
system~(TOF), and a CsI(Tl) electromagnetic calorimeter~(EMC), which are all
enclosed in a superconducting solenoidal magnet providing a 1.0~T magnetic
field. The solenoid is supported by an octagonal flux-return yoke with
resistive plate counter muon identifier modules interleaved with steel. The
acceptance of charged particles and photons is 93\% over a $4\pi$ solid angle.
The charged-particle momenta resolution at 1.0~GeV/$c$ is $0.5\%$, and the
specific energy loss~(d$E$/d$x$) resolution is $6\%$ for the electrons from
Bhabha scattering. The EMC measures photon energies with a resolution of
$2.5\%$~($5\%$) at $1$~GeV in the barrel (end-cap) region. The time resolution
of the TOF barrel part is 68~ps, while that of the end-cap part is 110~ps. The
end-cap TOF was  upgraded in 2015 with multi-gap resistive plate chamber
technology, providing a time resolution of 60~ps.

The data samples used in this analysis are listed in Table~\ref{energe}.
For some aspects of the analysis, these samples are organized into three
sample groups, 4.178~GeV, 4.189-4.219~GeV, and 4.226~GeV, that were acquired 
during the same year under consistent running conditions. Since the cross 
section of $D_{s}^{*\pm}D_{s}^{\mp}$ production in $e^{+}e^{-}$ annihilation is 
about a factor of twenty larger than that of $D_{s}^{+}D_{s}^{-}$~\cite{DsStrDs}, 
and the $D_{s}^{*\pm}$ meson decays to $\gamma D_{s}^{\pm}$ with a dominant BF 
of $(93.5\pm0.7)$\%~\cite{PDG}, the signal events discussed in this paper are 
selected from the process $e^+e^-\to D_{s}^{*\pm}D_{s}^{\mp}\to \gamma D_{s}^{+}D_{s}^{-}$.

 \begin{table}[htb]
 \renewcommand\arraystretch{1.25}
 \centering
 \caption{The integrated luminosities ($\mathcal{L}_{\rm int}$) and the 
   requirements on $M_{\rm rec}$ for various collision energies. The definition 
   of $M_{\rm rec}$ is given in Eq.~(\ref{eq:mrec}). The first and second 
   uncertainties are statistical and systematic, respectively.}
 \begin{tabular}{ccc}
 \hline
 $\sqrt{s}$ (GeV) & $\mathcal{L}_{\rm int}$ (pb$^{-1}$) & $M_{\rm rec}$ (GeV/$c^2$)\\
 \hline
  4.178 &3189.0$\pm$0.2$\pm$31.9&[2.050, 2.180] \\
  4.189 &526.7$\pm$0.1$\pm$2.2&[2.048, 2.190] \\
  4.199 &526.0$\pm$0.1$\pm$2.1&[2.046, 2.200] \\
  4.209 &517.1$\pm$0.1$\pm$1.8&[2.044, 2.210] \\
  4.219 &514.6$\pm$0.1$\pm$1.8&[2.042, 2.220] \\
  4.226 &1047.3$\pm$0.1$\pm10.2$&[2.040, 2.220] \\
  \hline
 \end{tabular}
 \label{energe}
\end{table}

Simulated samples are produced with the {\sc geant4}-based~\cite{GEANT4}
Monte Carlo~(MC) package, which includes the geometric description of the BESIII
detector and the detector response. These samples are used to determine the detection
efficiency and to estimate the background. The simulation includes the beam-energy 
spread and initial-state radiation~(ISR) in  $e^+e^-$ annihilations
modeled with the generator {\sc kkmc}~\cite{KKMC}. The generic MC samples
consist of the production of $D\bar{D}$ pairs with consideration of quantum
coherence for all neutral $D$ modes, the non-$D\bar{D}$ decays of the $\psi(3770)$,
the ISR production of the $J/\psi$ and $\psi(3686)$ states, and the continuum
processes. The known decay modes are modeled with {\sc evtgen}~\cite{EVTGEN1, EVTGEN2}
using the BFs taken from the Particle Data Group~(PDG)~\cite{PDG}, and the
remaining unknown decays from the charmonium states with
{\sc lundcharm}~\cite{LUNDCHARM1, LUNDCHARM2}. Final-state radiation from charged particles is
incorporated with the {\sc photos}~\cite{PHOTOS} package.

\section{Event selection}
\label{ST-selection}
The data samples were collected just above the $D_s^{*\pm}D_s^{\mp}$ threshold. 
The tag method allows clean signal samples to be selected, which provide an 
opportunity to perform amplitude analyses and to measure the absolute BFs of 
the hadronic $D^+_s$ meson decays. In the tag method, a single-tag~(ST) 
candidate requires only one of the $D_{s}^{\pm}$ mesons to be reconstructed via 
a hadronic decay; a double-tag~(DT) candidate has both $D_s^+D_s^-$ mesons 
reconstructed via hadronic decays. The DT candidates are required to have the 
$D_{s}^{+}$ meson decaying to the signal mode 
$D_{s}^{+} \to K^0_{S}\pi^{+}\pi^{0}$ and the $D_{s}^{-}$ meson decaying to a tag 
mode. (Charge conjugation is implied throughout this paper.) Nine tag modes are 
reconstructed and the corresponding mass windows on the tagging $D_{s}^{-}$ 
mass~($M_{\rm tag}$) are listed in Table~\ref{tab:tag-cut}. The $D_{s}^{\pm}$ 
candidates are constructed from individual $\pi^\pm$, $K^\pm$, $\eta$, 
$\eta^{\prime}$, $K_{S}^{0}$ and $\pi^{0}$ particles.

\begin{table}[htbp]
 \renewcommand\arraystretch{1.25}
 \centering
\caption{Requirements on $M_{\rm tag}$ for various tag modes, where the
  $\eta$ and $\eta^\prime$ subscripts denote the decay modes used to reconstruct 
  these particles.}\label{tab:tag-cut}
     \begin{tabular}{lc}
        \hline
        Tag mode                                     & Mass window (GeV/$c^{2}$) \\
        \hline
        $D_{s}^{-} \to K_{S}^{0}K^{-}$               & [1.948, 1.991]            \\
        $D_{s}^{-} \to K^{+}K^{-}\pi^{-}$            & [1.950, 1.986]            \\
        $D_{s}^{-} \to K_{S}^{0}K^{+}\pi^{0}$        & [1.946, 1.987]            \\
        $D_{s}^{-} \to K^{+}K^{-}\pi^{-}\pi^{0}$     & [1.947, 1.982]            \\
        $D_{s}^{-} \to K_{S}^{0}K^{+}\pi^{-}\pi^{-}$ & [1.953, 1.983]            \\
        $D_{s}^{-} \to \pi^{-}\pi^{-}\pi^{+}$        & [1.952, 1.982]            \\
        $D_{s}^{-} \to \pi^{-}\eta_{\gamma\gamma}$   & [1.930, 2.000]            \\
        $D_{s}^{-} \to \pi^{-}\pi^{0}\eta_{\gamma\gamma}$   & [1.920, 2.000]     \\
        $D_{s}^{-} \to \pi^{-}\eta_{\pi^{+}\pi^{-}\eta_{\gamma\gamma}}^{\prime}$
                                                     & [1.940, 1.996]            \\
        \hline
      \end{tabular}
\end{table}

Charged track candidates from the MDC must satisfy $|\!\cos\theta|<0.93$,
where $\theta$ is the polar angle with respect to the direction of the positron
beam. The closest approach to the interaction point is required to be less
than 10~cm along the beam direction and less than 1 cm in the plane perpendicular
to the beam. Particle identification (PID) of charged particles is implemented
by combining the d$E$/d$x$ information in the MDC and the time-of-flight information
from the TOF system. For charged kaon (pion) candidates, the probability for the
kaon (pion) hypothesis is required to be larger than that for the pion (kaon)
hypothesis.

The $K_{S}^{0}$ mesons are reconstructed with pairs of two oppositely
charged tracks, which satisfy $|\!\cos\theta|< 0.93$ and the distances of
closest approach along the beam direction must be less than 20~cm.
The decay length of the reconstructed $K_{S}^{0}$ in the signal side decay is
required to be more than twice that of the vertex resolution away from the
interaction point. The invariant masses of these charged track pairs are
required to be in the range $[0.487, 0.511]$~GeV/$c^{2}$.

Photons are reconstructed from the clusters of deposited energy in the EMC. The shower time
is required to be within $[0, 700]$~ns of the event start time in order to
suppress electronics noise or $e^+e^-$ beam background. Photon candidates within
$|\!\cos\theta| < 0.80$ (barrel) are required to have an energy deposition
larger than $25$~MeV and those with $0.86<|\!\cos\theta|<0.92$ (end-cap) must
have an energy deposition larger than $50$~MeV. To suppress the noise from
hadronic shower splitoffs, the calorimeter positions of photon candidates must
lie outside a cone of $10\degree$ from all charged tracks.
The $\pi^0$ $(\eta)$ candidates are reconstructed through
$\pi^0\to \gamma\gamma$ ($\eta \to \gamma\gamma$) decays,
with at least one barrel photon. The invariant mass of the photon pair for
$\pi^{0}$ and $\eta$ candidates must be in the ranges $[0.115, 0.150]$~GeV/$c^{2}$
and $[0.490, 0.580]$~GeV/$c^{2}$, respectively.
A kinematic fit that constrains the $\gamma\gamma$ invariant mass to the $\pi^{0}$
or $\eta$ nominal mass~\cite{PDG} is performed to improve the mass resolution.
The $\chi^2$ of the kinematic fit is required to be less than 30.
The $\eta^{\prime}$ candidates are formed from the $\pi^{+}\pi^{-}\eta$
combinations with an invariant mass within a range of
$[0.946, 0.970]$~GeV/$c^{2}$.

$D_{s}^{\pm}$ candidates with $M_{\rm rec}$ lying with the mass windows listed in Table~\ref{energe} are retained for further
study. The quantity $M_{\rm rec}$ is defined as
\begin{eqnarray}
\begin{aligned}
    M_{\rm rec} = \sqrt{(E_{\rm cm} - \sqrt{|\vec{p}_{D_{s}}|^{2}+m_{D_{s}}^{2}})^{2} - |\vec{p}_{D_{s}} | ^{2}} \; , \label{eq:mrec}
\end{aligned}\end{eqnarray}
where $E_{\rm cm}$ is the initial energy of the $e^+e^-$ center-of-mass system,
$\vec{p}_{D_{s}}$ is the three-momentum of the $D_{s}^{\pm}$ candidate in the
$e^+e^-$ center-of-mass frame, and $m_{D_{s}}$ is the $D_{s}^{\pm}$ nominal mass~\cite{PDG}.

A ``$K_{S}^{0}K$'' veto and a ``$D^{0}$'' veto are applied on the signal
$D_{s}^{+}$ candidates. The Cabibbo-favored $D_{s}^{+} \to K_{S}^{0}K^{+}$
decay contributes background when the $K^{+}$ is misreconstructed as a $\pi^{+}$.
This background is reduced by a veto on the signal $D_{s}^{+}$, with
$M_{K_{S}^{0}K^{+}}-m_{D_{s}}< -20$~MeV/$c^{2}$, where $M_{K_{S}^{0}K^{+}}$ is the
invariant mass of the $K_{S}^{0}$ and reconstructed $\pi^{+}$ track, when it is assumed to
be a kaon.
There is also track-swap background where $D^{0} \to K_{S}^{0}K^{-}\pi^{+}$
versus~$\bar{D}^{0} \to K^{+}\pi^{-}\pi^{0}$  fake
$D_{s}^{+} \to K_{S}^{0}\pi^+\pi^0$ versus~$D_{s}^{-} \to K^{+}K^{-}\pi^-$ events through the
exchange of $K_{S}^{0}$ and $K^{+}$, or $\pi^{0}$ and $K^{-}$ tracks. Events which
simultaneously satisfy $|M_{K_{S}^{0}K^{-}\pi^{+}} -m_{D^{0}}|<30$~MeV/$c^{2}$ and
$|M_{K^{+}\pi^{-}\pi^{0}}-m_{D^{0}}|<30$~MeV/$c^{2}$ are rejected, where
$M_{K_{S}^{0}K^{-}\pi^{+}}$ ($M_{K^{+}\pi^{-}\pi^{0}}$) is the invariant mass
of the $K_{S}^{0}K^{-}\pi^{+}$ ($K^{+}\pi^{-}\pi^{0}$) combination and $m_{D^{0}}$ is the $D^{0}$
nominal mass~\cite{PDG}.

\section{Amplitude analysis of $D_{s}^{+} \to K^0_{S}\pi^{+}\pi^{0}$}
\label{Amplitude-Analysis}
\subsection{Event selection}
\label{AASelection}
An eight-constraint kinematic fit is performed assuming the process
$e^+e^-\to D_{s}^{*\pm}D_{s}^{\mp}\to \gamma D_{s}^{+}D_{s}^{-}$,
with $D_{s}^{-}$ decaying to one of the tag modes and $D_{s}^{+}$ decaying to the
signal mode. The combination with the  minimum $\chi^2$ is chosen, assuming that a $D_s^{*+}$
meson decays to $D_{s}^{+}\gamma$ or a $D_s^{*-}$ meson decays to
$D_{s}^{-}\gamma$. In addition to the constraints of four-momentum
conservation in the $e^+e^-$ center-of-mass system, the invariant masses of $(\gamma\gamma)_{\pi^{0}}$,
$(\pi^{+}\pi^{-})_{K_S^0}$, tag $D_{s}^{-}$, and $D_{s}^{*\pm}$ candidates are
constrained to the corresponding nominal masses~\cite{PDG}.
In order to ensure that all candidates fall within the phase-space boundary,
the constraint of the signal $D_s^{+}$ mass is added to the kinematic fit
and the updated four-momenta are used for the amplitude analysis.

\begin{figure*}[!htbp]
  \centering
  \includegraphics[width=0.32\textwidth]{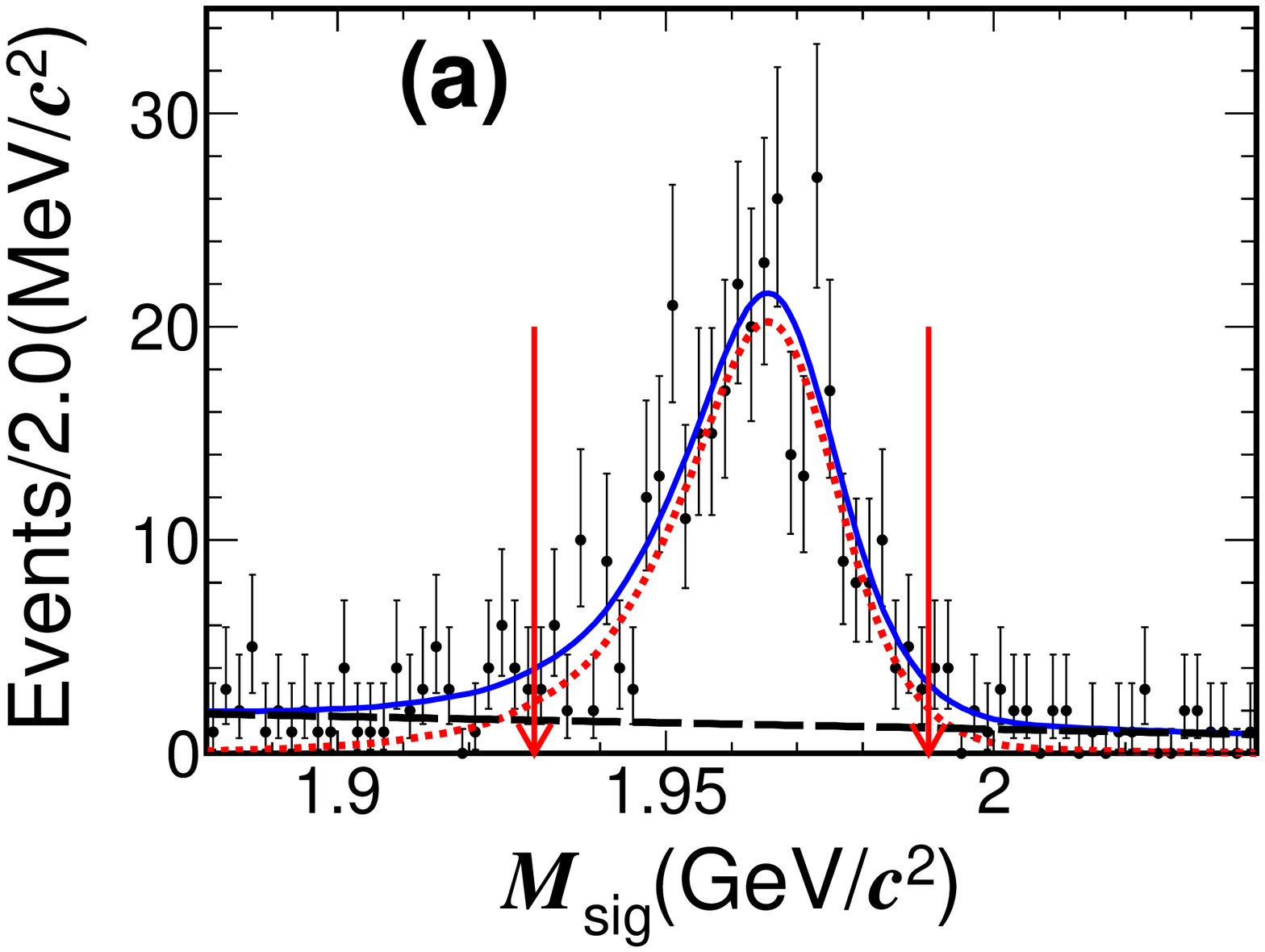}
  \includegraphics[width=0.32\textwidth]{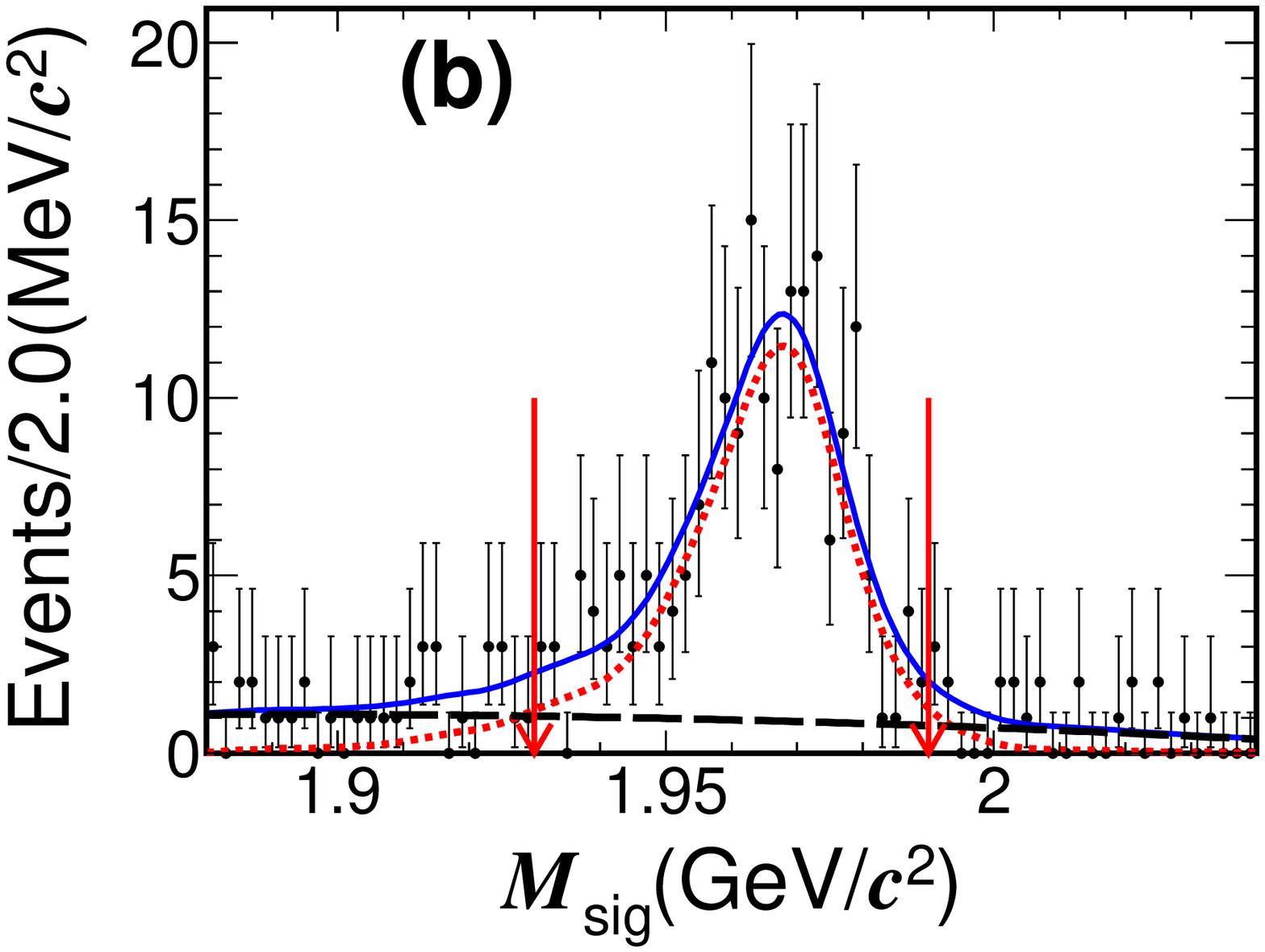}
  \includegraphics[width=0.32\textwidth]{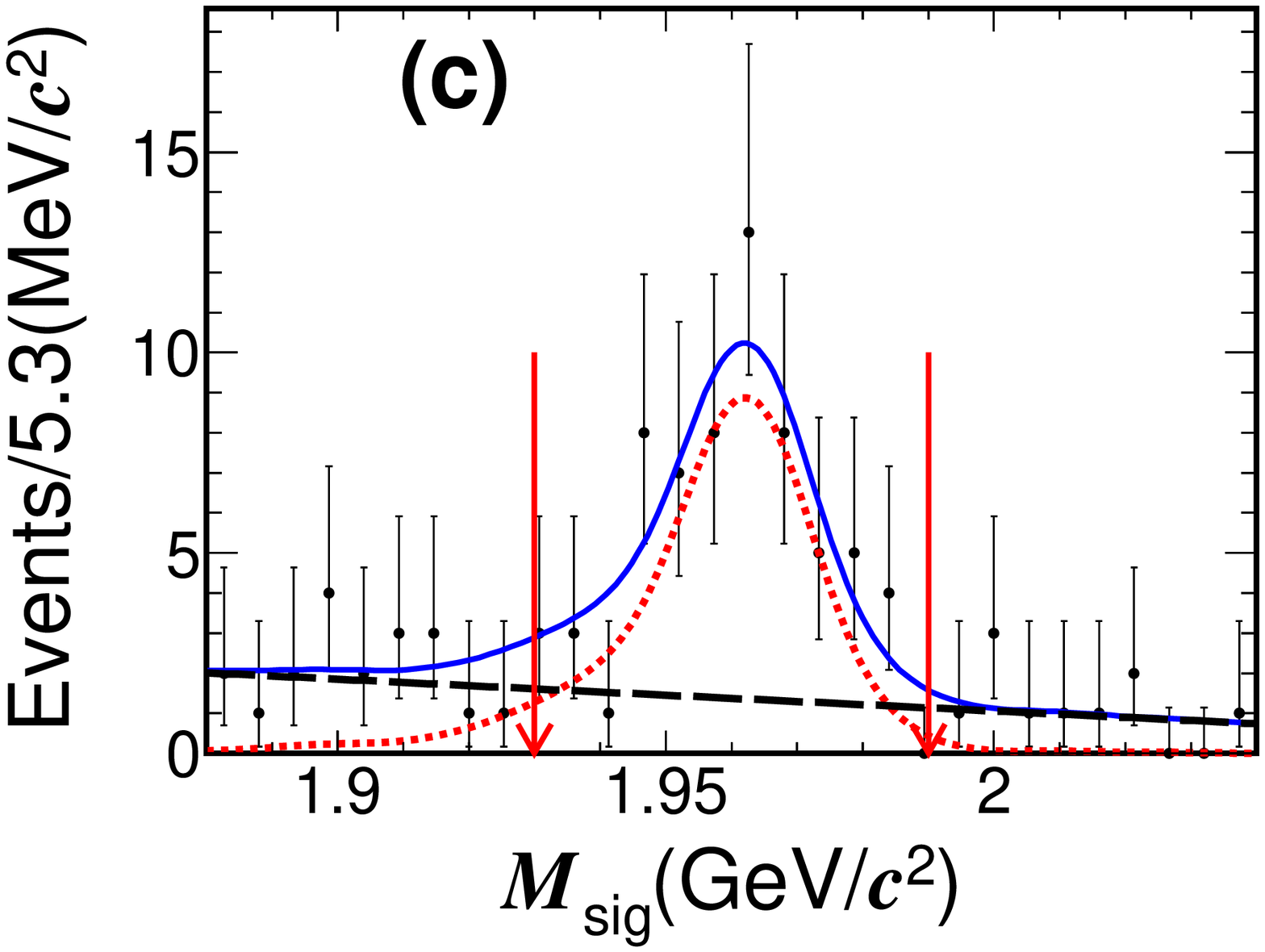}
  \caption{
    Fits to the $M_{\rm sig}$ distributions of the data samples at $\sqrt{s}=$ (a) 4.178~GeV,
    (b) 4.189-4.219~GeV, and (c) 4.226~GeV. The black points with error
    bars are data. The blue solid lines are the total fits. The red dotted and the black
    dashed lines are the fitted signal and background, respectively. The pairs of red
    arrows indicate the signal regions.
  } \label{fig:fit_Ds}
\end{figure*}

Furthermore, it is required that the energy of the transition photon from
$D_s^{*\pm}\to \gamma D_s^{\pm}$ is smaller than 0.18~GeV and the
mass recoiling against this photon and the signal $D_s^{+}$ candidate lies
within the range $[1.952, 1.995]$~GeV/$c^2$. Finally, a mass window,
$[1.930, 1.990]$~GeV/$c^2$, is applied on the signal $D_s^{+}$ candidates.
Fig.~\ref{fig:fit_Ds} shows the fits to the invariant-mass distributions of
the accepted signal $D_s^{+}$ candidates, $M_{\rm sig}$, for various data
samples. In the fits, the signal is described by an MC-simulated shape
convolved with a Gaussian resolution function, and the background is described by a
second-order Chebyshev function. 
There are $352$, $193$, and $64$ events retained for
the amplitude analysis with purities, $w_{\rm sig}$, of $88.9\pm6.8\%$, $84.6\pm8.3\%$, and
$75.9\pm14.3\%$ for the data samples at $\sqrt{s}=4.178$~GeV, 4.189-4.219~GeV, and 4.226~GeV,
respectively.

\subsection{Fit method}
The intermediate-resonance composition in the decay $D_{s}^{+}\to K_{S}^{0}\pi^+\pi^0$
is determined by an unbinned maximum-likelihood fit to data. The likelihood
function is constructed with a probability density function (PDF), which depends
on the momenta of the three daughter particles. The amplitude of the $n$th
intermediate state~($A_{n}$) is
\begin{eqnarray}
\begin{aligned}
  A_{n} = P_{n}S_{n}F_{n}^{r}F_{n}^{D}\,, \label{base-amplitude}
\end{aligned}\end{eqnarray}
where $S_{n}$ and $F_{n}^{r(D)}$ are the spin factor and the Blatt-Weisskopf
barriers of the intermediate state (the $D_{s}^{\pm}$ meson), respectively, and
$P_{n}$ is the propagator of the intermediate resonance.

The total amplitude $M$ is then the coherent sum of the amplitudes of
intermediate processes, $M=\begin{matrix}\sum c_{n}A_{n}\end{matrix}$,
where $c_{n}=\rho_{n}e^{i\phi_{n}}$ is the corresponding complex coefficient.
The magnitude $\rho_{n}$ and phase $\phi_{n}$ are free parameters in the fit,
and are defined relative to those of a reference mode, for which they are fixed.
The signal PDF $f_{S}(p_{j})$ is given by
\begin{eqnarray}\begin{aligned}
  f_{S}(p_{j}) = \frac{\epsilon(p_{j})\left|M(p_{j})\right|^{2}R_{3}(p_{j})}{\int \epsilon(p_{j})\left|M(p_{j})\right|^{2}R_{3}(p_{j})\,dp_{j}}\,, \label{signal-PDF}
\end{aligned}\end{eqnarray}
where $\epsilon(p_{j})$ is the detection efficiency parameterized in terms of
the final four-momenta $p_{j}$. The index $j$ refers to the different particles
in the final states, and $R_{3}(p_{j})$ is the standard element of three-body
phase space. The normalization integral is determined by an MC integration,
\begin{eqnarray}\begin{aligned}
  \int \epsilon(p_{j})\left|M(p_{j})\right|^{2}R_{3}(p_{j})\,dp_{j} \approx
\frac{1}{N_{M}}\sum_{k}^{N_{M}} \frac{\left|M(p_{j}^{k})\right|^{2}}{\left|M^{g}(p_{j}^{k})\right|^{2}}\,, \label{MC-intergral}
\end{aligned}\end{eqnarray}
where $k$ is the index of the $k$th event and $N_{M}$ is the number of the
selected MC events. Here $M^{g}(p_{j})$ is the PDF used to generate the MC
samples in MC integration. To account for any bias caused by differences in
PID and tracking  efficiency  between data and MC simulation, each signal MC event is
weighted with a ratio, $\gamma_{\epsilon}(p)$, of the efficiency of data to
that of MC simulation and the MC integration then becomes
\begin{eqnarray}\begin{aligned}
    &\int \epsilon(p_{j})\left|M(p_{j})\right|^{2}R_{3}(p_{j})\,dp_{j} \approx
&\frac{1}{N_{M}} \sum_{k}^{N_{M}} \frac{\left|M(p_{j}^{k})\right|^{2}\gamma_{\epsilon}(p_{j}^{k})}{\left|M^{g}(p_{j}^{k})\right|^{2}}\,.
\label{MC-intergral-corrected}
\end{aligned}\end{eqnarray}
Finally, the log-likelihood is written as
\begin{eqnarray}\begin{aligned}
  \ln{\mathcal{L}} = \sum_{k}^{N_{D}} \ln f_{S}(p_{j}^{k})\,,  \label{likelihood}
\end{aligned}\end{eqnarray}
where $N_{D}$ is the number of candidate events in data.

Equation~(\ref{likelihood}), however,  is only appropriate for a data sample with
negligible background. A signal-background combined PDF is introduced to
account for the approximate 15\% of background in this analysis. The background PDF
is given by
\begin{eqnarray}\begin{aligned}
  f_{B}(p_{j}) = \frac{B(p_{j})R_{3}(p_{j})}{\int B(p_{j})R_{3}(p_{j})\,dp_{j}}\,.\label{bkg-PDF}
\end{aligned}\end{eqnarray}
The background shape $B(p_{j})$ is derived from about 3500 background
events selected from generic MC samples using RooNDKeysPdf~\cite{Verkerke} with the Dalitz
distribution of $M^2_{K_S^0\pi^+}$ versus~$M^2_{K_S^0\pi^0}$ as input.
This background PDF is added to the signal PDF incoherently. Then, the combined
PDF is written as
\begin{eqnarray}
\begin{aligned}
  w_{\rm sig}f_{S}(p_{j})&+(1-w_{\rm sig})f_{B}(p_{j})\\
  &=w_{\rm sig}\frac{\epsilon(p_{j})\left|M(p_{j})\right|^{2}R_{3}(p_{j})}{\int \epsilon(p_{j})\left|M(p_{j})\right|^{2}R_{3}(p_{j})\,dp_{j}}\\
  &+(1-w_{\rm sig})\frac{B(p_{j})R_{3}(p_{j})}{\int B(p_{j})R_{3}(p_{j})\,dp_{j}}\,. \label{combined-PDF}
\end{aligned}
\end{eqnarray}
Thus, $\epsilon(p_{j})$ is factorized from the background PDF to make it a
comparable equation with the signal PDF. In this way, the $\epsilon(p_{j})$ term,
which is independent of the fitted variables, is regarded as a constant and can be
dropped during the log-likelihood fit. The background shape is extracted from the
selected events, hence one has to manually divide the background PDF by the
efficiency, $B_{\epsilon}\equiv B/\epsilon$. As a consequence, the combined PDF
becomes
\begin{eqnarray}
\begin{aligned}
  w_{\rm sig}f_{S}(p_{j})+(1-w_{\rm sig})&f_{B}(p_{j})\\
  =\epsilon(p_{j})R_{3}(p_{j})&\left[\frac{w_{\rm sig}\left|M(p_{j})\right|^{2}}{\int \epsilon(p_{j})\left|M(p_{j})\right|^{2}R_{3}(p_{j})\,dp_{j}}\right.\\
  &+\left.\frac{(1-w_{\rm sig})B_{\epsilon}(p_{j})}{\int \epsilon(p_{j})B_{\epsilon}(p_{j})R_{3}(p_{j})\,dp_{j}}
\right]\,. \label{combined-PDF-2}
\end{aligned}
\end{eqnarray}
Next, the integration in the denominator of the background term can also be
handled by the MC integration method in the same way as for the signal only sample:
\begin{eqnarray}\begin{aligned}
  \int \epsilon(p_{j})B_{\epsilon}(p_{j})R_{3}(p_{j})\,dp_{j} \approx
\frac{1}{N_{M}}\sum_{k}^{N_{M}} \frac{B_{\epsilon}(p_{j}^{k})}{\left|M^{g}(p_{j}^{k})\right|^{2}}\,.
\end{aligned}\end{eqnarray}
Eventually, for the case with a  non-negligible background, Eq.~(\ref{likelihood}) is extended
to become
\begin{eqnarray}\begin{aligned}
  \ln{\mathcal{L}} = \sum_{i=1}^{3}\sum_{k}^{N_{D, i}} \ln\left[w_{\rm sig}^{i}f_{S}(p_{j}^{k})+(1-w_{\rm sig}^{i})f_{B}(p_{j})\right]\,,  \label{likelihood3}
\end{aligned}\end{eqnarray}
where $i$ indicate the data sample and $w_{\rm sig}^{i}$ are fixed during
the log-likelihood fit.

\subsubsection{Blatt-Weisskopf barrier factors}
For the process $a \to bc$, the Blatt-Weisskopf barrier $F_L(p_j)$ is
parameterized as a function of the angular momenta $L$ and the momenta $q$ of
the daughter $b$ or $c$ in the rest system of $a$,
\begin{eqnarray}
\begin{aligned}
 F_{L=0}(q)&=1,\\
 F_{L=1}(q)&=\sqrt{\frac{z_0^2+1}{z^2+1}},\\
 F_{L=2}(q)&=\sqrt{\frac{z_0^4+3z_0^2+9}{z^4+3z^2+9}}\,,
\end{aligned}
\end{eqnarray}
where $z=qR$ and $z_0=q_0R$. The effective radius of the barrier $R$ is fixed to
3.0~GeV$^{-1}$ for the intermediate resonances and 5.0~GeV$^{-1}$ for the $D_s^+$
meson.

\subsubsection{Propagator}
The intermediate resonances $K^{*}(892)^{0, +}$ and $K^{*}(1410)^{0}$
are parameterized as relativistic Breit-Wigner
functions,
\begin{eqnarray}\begin{aligned}
  \begin{array}{lr}
    P = \frac{1}{(m_{0}^{2} - s_{a} ) - im_{0}\Gamma(m)}\,, &\\
    \Gamma(m) = \Gamma_{0}\left(\frac{q}{q_{0}}\right)^{2L+1}\left(\frac{m_{0}}{m}\right)\left(\frac{F_{L}(q)}{F_{L}(q_{0})}\right)^{2}\,, &
  \end{array}\label{RBW}
\end{aligned}\end{eqnarray}
where $s_{a}$ denotes the invariant-mass squared of the parent particle;
$m_{0}$ and $\Gamma_{0}$ are the rest masses and the widths of the intermediate
resonances, respectively, and are fixed to the PDG values~\cite{PDG}.

We parameterize the $\rho^{+}$ and $\rho(1450)^+$ resonances by the Gounaris-Sakurai
lineshape~\cite{PhysRevLett.21.244}, which is given by
\begin{eqnarray}\begin{aligned}
P_{\rm GS}(m)=\frac{1+d\frac{\Gamma_0}{m_0}}{(m_0^2-m^2)+f(m)-im_0\Gamma(m)}\,.
\end{aligned}\end{eqnarray}
The function $f(m)$ is given by
\begin{eqnarray}
\begin{aligned}
f(m)&=\Gamma_0\frac{m_0^2}{q_0^3}\left[q^2(h(m)-h(m_0))\phantom{\left.\frac{d h}{d(m^2)}\right|_{m_0^2}}\right.\\
    &+\left.(m_0^2-m^2)q_0^2\left.\frac{dh}{d(m^2)}\right|_{m_0^2}\right]\,,
\end{aligned}
\end{eqnarray}
where
\begin{eqnarray}\begin{aligned}
h(m)=\frac{2q}{\pi m}\ln\left(\frac{m+2q}{2m_{\pi}}\right)\,,
\end{aligned}\end{eqnarray}
and
\begin{eqnarray}\begin{aligned}
&\left.\frac{dh}{d(m^2)}\right|_{m_0^2}=
&h(m_0)\left[(8q_0^2)^{-1}-(2m_0^2)^{-1}\right]+(2\pi m_0^2)^{-1}\,.
\end{aligned}\end{eqnarray}
The normalization condition at $P_{\rm GS}(0)$ fixes the parameter
$d=f(0)/(\Gamma_0 m_0)$. It is found to be
\begin{eqnarray}\begin{aligned}
d=\frac{3m^2_\pi}{\pi q_0^2}\ln\left(\frac{m_0+2q_0}{2m_\pi}\right)+\frac{m_0}{2\pi q_0}-\frac{m^2_\pi m_0}{\pi q^3_0}\,.
\end{aligned}\end{eqnarray}

\subsubsection{Spin factors}
The spin-projection operators are defined as~\cite{covariant-tensors}
\begin{eqnarray}
\begin{aligned}
  P^{(1)}_{\mu\mu^{\prime}}(a) &= -g_{\mu\mu^{\prime}}+\frac{p_{a,\mu}p_{a,\mu^{\prime}}}{p_{a}^{2}}\,,\\
  P^{(2)}_{\mu\nu\mu^{\prime}\nu^{\prime}}(a) &= \frac{1}{2}(P^{(1)}_{\mu\mu^{\prime}}(a)P^{(1)}_{\nu\nu^{\prime}}(a)+P^{(1)}_{\mu\nu^{\prime}}(a)P^{(1)}_{\nu\mu^{\prime}}(a))\\
  &-\frac{1}{3}P^{(1)}_{\mu\nu}(a)P^{(1)}_{\mu^{\prime}\nu^{\prime}}(a)\,.
 \label{spin-projection-operators}
\end{aligned}
\end{eqnarray}
The quantities $p_a$, $p_b$, and $p_c$ are the momenta of particles $a$,
$b$, and $c$, respectively, and $r_a = p_b-p_c$.
The covariant tensors are given by
\begin{eqnarray}
\begin{aligned}
    \tilde{t}^{(1)}_{\mu}(a) &= -P^{(1)}_{\mu\mu^{\prime}}(a)r^{\mu^{\prime}}_{a}\,,\\
    \tilde{t}^{(2)}_{\mu\nu}(a) &= P^{(2)}_{\mu\nu\mu^{\prime}\nu^{\prime}}(a)r^{\mu{\prime}}_{a}r^{\nu^{\prime}}_{a}\,.\\
\label{covariant-tensors}
\end{aligned}
\end{eqnarray}
The spin factors for $S$, $P$, and $D$ wave decays are
\begin{eqnarray}
\begin{aligned}
    S &= 1\,, &(S\ \text{wave}), &\\
    S &= \tilde{T}^{(1)\mu}(D_{s}^{\pm})\tilde{t}^{(1)}_{\mu}(a)\,,         &(P\ \text{wave}),\\
    S &= \tilde{T}^{(2)\mu\nu}(D_{s}^{\pm})\tilde{t}^{(2)}_{\mu\nu}(a)\,,         &(D\ \text{wave}),
\label{spin-factor}
\end{aligned}
\end{eqnarray}
where the $\tilde{T}^{(l)}$ factors have the same definition as $\tilde{t}^{(l)}$. The
tensor describing the $D_{s}^{+}$ decay is denoted by $\tilde{T}$ and
that of the $a$ decay is denoted by $\tilde{t}$.

\subsection{Fit results}
\begin{figure}[htbp]
  \centering
  \mbox{
    \begin{overpic}[width=0.5\textwidth]{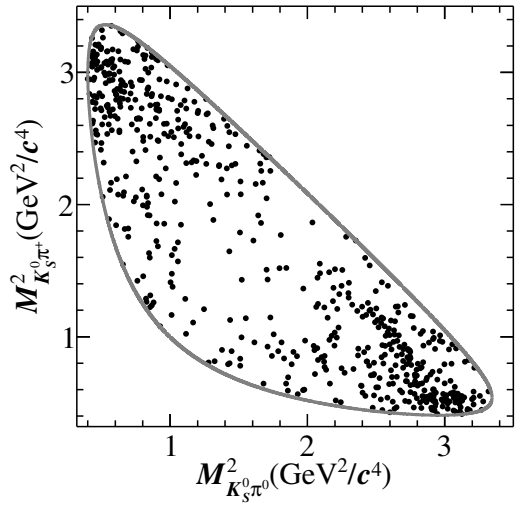}
    \end{overpic}
  }
  \caption{The Dalitz plot of the DT candidates from the data samples at
    $\sqrt{s}= 4.178$-$4.226$~GeV.}
  \label{dalitz}
\end{figure}

The Dalitz plot of $M^{2}_{K_S^0\pi^0}$ versus~$M^{2}_{K_S^0\pi^{+}}$ summed over all the
data samples is shown in Fig.~\ref{dalitz}. We can see an anti-diagonal band
corresponding to $K_{S}^{0}\rho^+$. In the fit, the magnitude and phase of the
reference amplitude $D_{s}^{+} \to K_S^0\rho^+$ are fixed to 1.0 and 0.0,
respectively, and the masses and widths of all resonances are fixed to the
corresponding PDG averages~\cite{PDG}. In addition to the dominating amplitude
$D_{s}^{+} \to K_S^0\rho^+$, we have tested for the contribution of all possible intermediate
resonances including $K^{*}(892)^{0}$, $K^{*}(892)^{+}$, $K^*(1410)$,
$K^*_0(1430)$, $K^*_2(1430)$, $\rho(1450)$, $K^*(1680)$, $\rho(1700)$, etc.
We find that $D^+_s\to K_{S}^{0}\rho(1450)^{+}$,
$D^+_s\to K^{*+}(892)\pi^0$, and $K^{*}(1410)^0\pi^+$ have a statistical
significances greater than three standard deviations and retain these amplitudes in the final model.

The calculation of the fit fractions (FFs) for individual amplitudes, involves the
phase-space MC truth information without  detector acceptance or resolution effects.
The FF for the $n$th amplitude is defined as
\begin{eqnarray}\begin{aligned}
  {\rm FF}_{n} = \frac{\sum^{N_{\rm gen}} \left|c_{n}A_{n}\right|^{2}}{\sum^{N_{\rm gen}} \left|M\right|^{2}}\,, \label{Fit-Fraction-Definition}
\end{aligned}\end{eqnarray}
where $N_{\rm gen}$ is the number of phase-space MC events at generator
level. These FFs will not sum to unity if there is net constructive or
destructive interference. Interference  IN between the $n$th and the
$n^{\prime}$th amplitudes is defined as (for $n<n^{\prime}$ only)
\begin{eqnarray}\begin{aligned}
  {\rm IN}_{nn^{\prime}} = \frac{\sum^{N_{\rm gen}} 2Re[c_{n}c^{*}_{n^{\prime}}A_{n}A^{*}_{n^{\prime}}]}{\sum^{N_{\rm gen}} \left|M\right|^{2}}\,. \label{interferenceFF-Definition}
\end{aligned}\end{eqnarray}
In order to determine the statistical uncertainties of FFs the amplitude
coefficients are
randomly sampled by a Gaussian-distributed amount set by the fit
uncertainty and the covariance matrix. Then the distribution of each FF is
fitted with a Gaussian function and the width of the Gaussian function is
defined as the uncertainty of the FF.

\begin{figure*}[!htbp]
 \centering
 \includegraphics[width=0.32\textwidth]{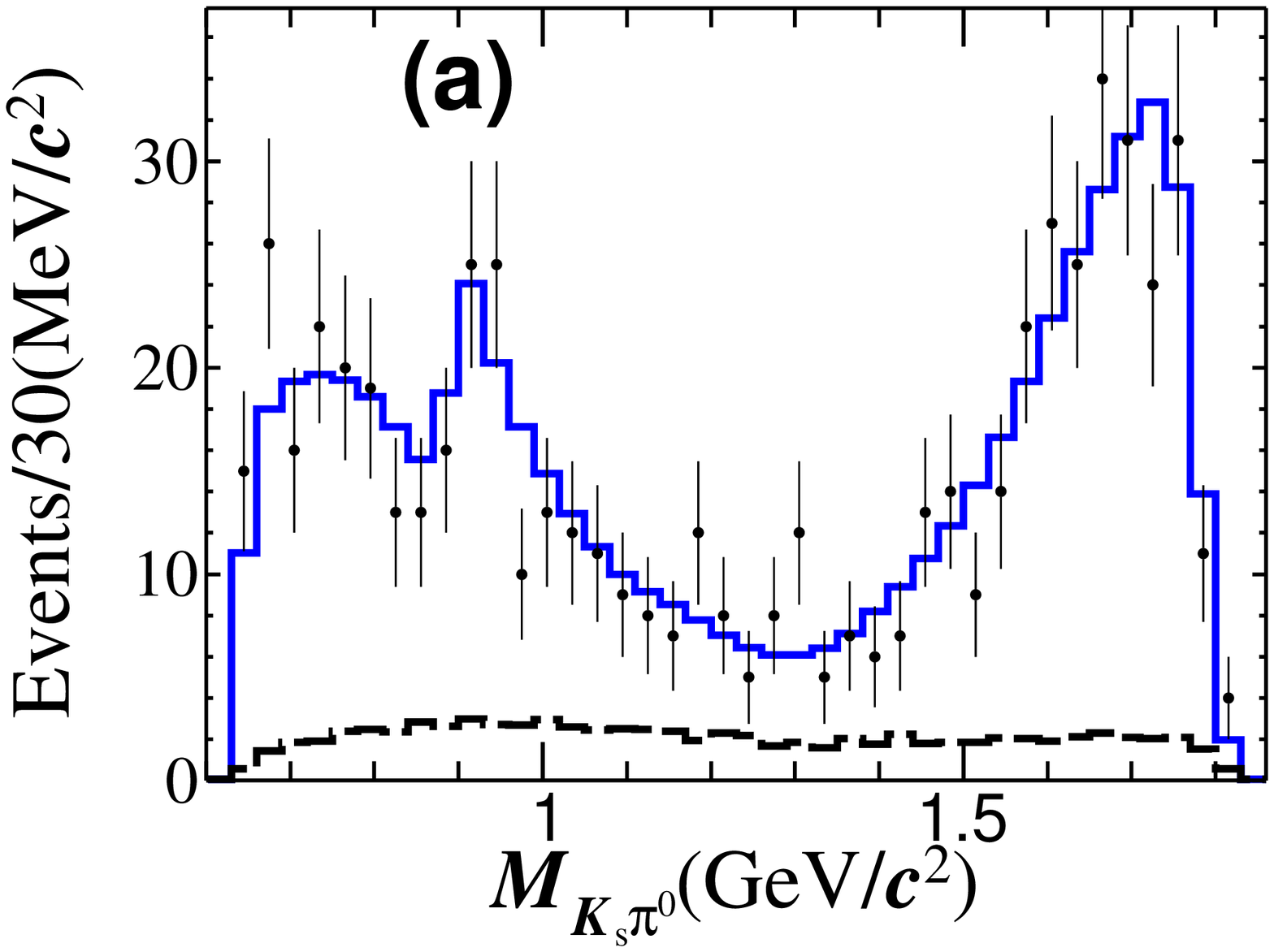}
 \includegraphics[width=0.32\textwidth]{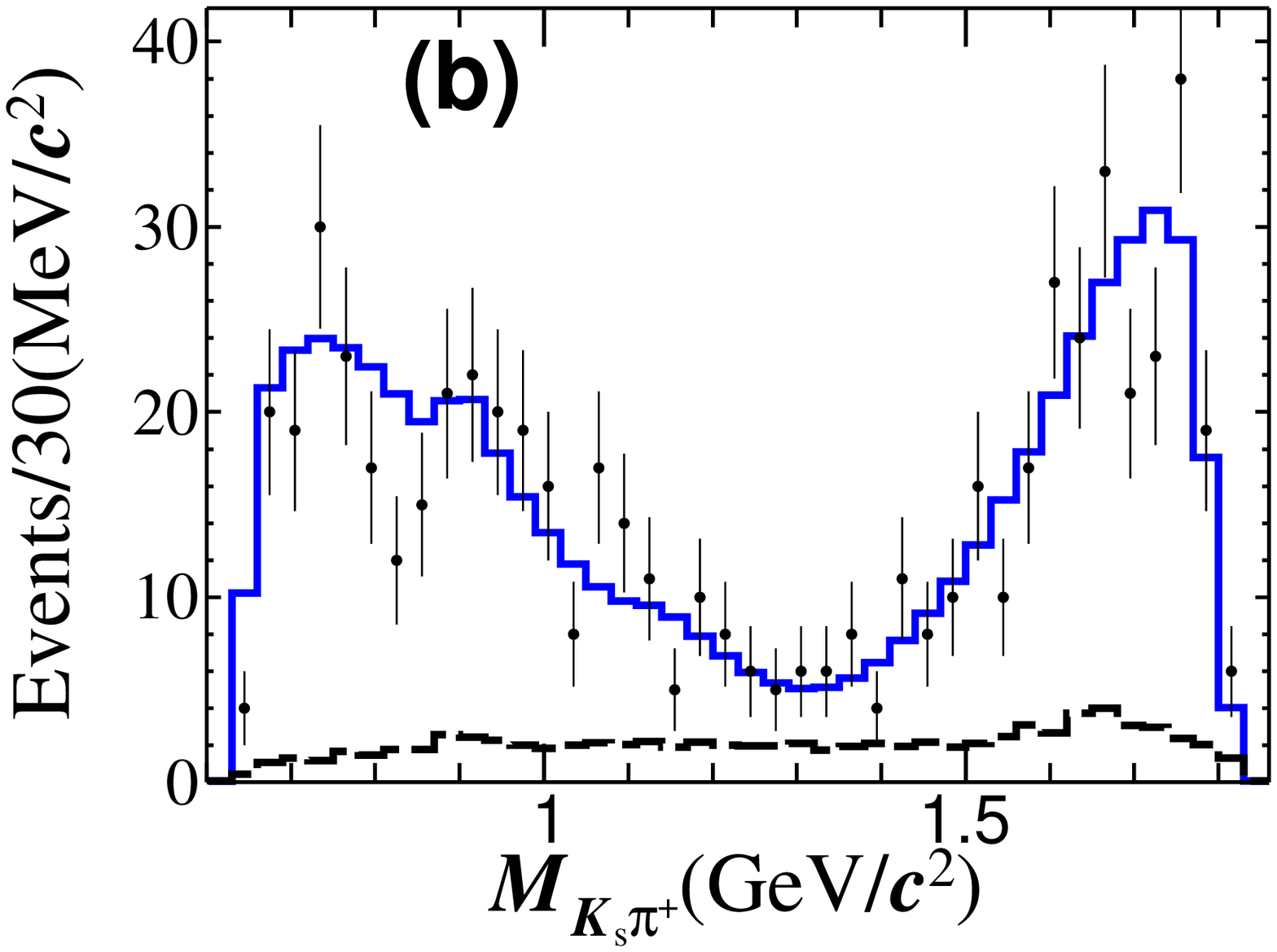}
 \includegraphics[width=0.32\textwidth]{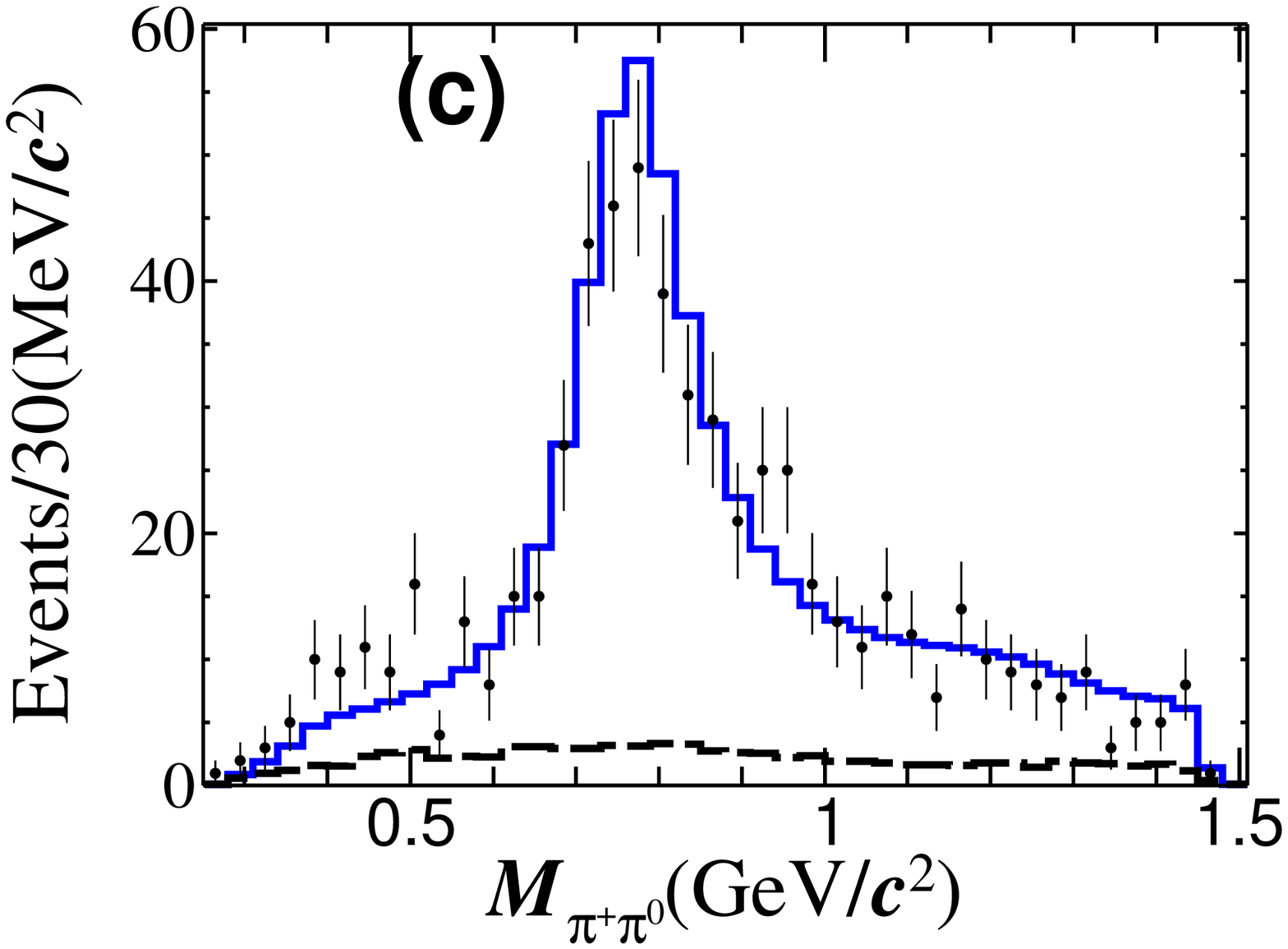}
 \caption{
   The projections of (a) $M_{K_S^0\pi^0}$, (b) $K_{S}^0\pi^+$, and (c)
   $M_{\pi^+\pi^0}$ from the nominal fit. The data samples at
   $\sqrt{s}= 4.178$-$4.226$~GeV are represented by points with error
   bars, the fit results by the solid blue lines, and the background
   estimated from generic MC samples by the black dashed lines.}
 \label{dalitz-projection}
\end{figure*}

The magnitudes, phases, FFs, and significances for the amplitudes are listed in
Table~\ref{fit-result}. The interference between amplitudes is listed in
Table~\ref{fit-interference}. The Dalitz plot projections are shown in
Fig.~\ref{dalitz-projection}. The assignment of  systematic uncertainties is
discussed in next section.

\begin{table*}[htbp]
    \caption{Magnitudes, phases, FFs, and significances for the amplitudes.
      The uncertainties in the magnitudes are statistical only. The first and the
      second uncertainties in the phases and FFs are statistical and systematic,
      respectively. The total FF is 86.9$\%$.}
    \label{fit-result}
    \begin{center}
    \begin{tabular}{lcccc}
      \hline
      Amplitude                               & Magnitude ($\rho_n$) & Phase ($\phi_n$)         & FF~(\%)                           &Significance~($\sigma$)\\
      \hline
      $D_{s}^{+} \to K_{S}^{0}\rho^{+}$       & 1.0(fixed)           & 0.0(fixed)               & $50.2 \pm 7.2 \pm 3.9$            &$>$10 \\
      $D_{s}^{+} \to K_{S}^{0}\rho(1450)^{+}$ & $2.7 \pm 0.5$        & $ 2.2 \pm 0.2 \pm 0.1$   & $20.4 \pm 4.3 \pm 4.4$            &$>$10\\
      $D_{s}^{+} \to K^{*}(892)^{0}\pi^{+}$   & $0.4 \pm 0.1$        & $ 3.2 \pm 0.2 \pm 0.1$   & $\phantom{0}8.4  \pm 2.2 \pm 0.9$ &5.0\\
      $D_{s}^{+} \to K^{*}(892)^{+}\pi^{0}$   & $0.3 \pm 0.1$        & $ 0.2 \pm 0.2 \pm 0.2$   & $\phantom{0}4.6  \pm 1.4 \pm 0.4$ &4.0\\
      $D_{s}^{+} \to K^{*}(1410)^{0}\pi^{+}$  & $0.8 \pm 0.2$        & $ 0.2 \pm 0.3 \pm 0.1$   & $\phantom{0}3.3  \pm 1.6 \pm 0.5$ &3.7\\
      \hline
    \end{tabular}
    \end{center}
\end{table*}
\begin{table*}[htbp]
    \caption{Interference between amplitudes, in  \% of the total amplitude.
      A denotes $D_{s}^{+} \to K_{S}^{0}\rho^{+}$,
      B $D_{s}^{+} \to K_{S}^{0}\rho(1450)^{+}$,
      C $D_{s}^{+} \to K^{*}(892)^{0}\pi^{+}$,
      D $D_{s}^{+} \to K^{*}(892)^{+}\pi^{0}$, and
      E $D_{s}^{+} \to K^{*}(1410)^{0}\pi^{+}$.
      The uncertainties are statistical only.}
    \label{fit-interference}
    \begin{center}
    \begin{tabular}{ccccc}
      \hline
                    & B              & C                         & D                         & E\\
    A                & 20.3 $\pm$ 5.3 & -4.1 $\pm$ 1.0            & -2.6 $\pm$ 0.9            & 5.1 $\pm$ 1.6\\
    B                &                & -4.5 $\pm$ 0.9            & -3.2 $\pm$ 0.7            & 0.8 $\pm$ 1.7\\
    C                &                &                           & -0.5 $\pm$ 0.1            & 0.4 $\pm$ 1.0\\
    D                &                &                           &                           & 0.5 $\pm$ 0.4 \\
    \hline
    \end{tabular}
    \end{center}
\end{table*}

\subsection{Systematic uncertainties for amplitude analysis}
\label{sec:PWA-Sys}
The systematic uncertainties for the amplitude analysis are summarized
in Table~\ref{systematic-uncertainties}, with their assignment described below.
\begin{itemize}
\item[\lowercase\expandafter{\romannumeral1}]
Resonance parameters. The masses and the widths of $\rho^{+}$, $\rho(1450)^{+}$,
$K^*(892)^{0(+)}$, and $K^*(1410)^{0}$ are shifted by their
corresponding uncertainties~\cite{PDG}.

\item[\lowercase\expandafter{\romannumeral2}]
R values. The radii of the nonresonant state and $D_s^{\pm}$ mesons
are varied within the range $[2.0, 4.0]$~GeV$^{-1}$ for intermediate
resonances and $[3.0, 7.0]$~GeV$^{-1}$ for $D_s^{\pm}$ mesons.

\item[\lowercase\expandafter{\romannumeral3}]
Background estimation.
The uncertainties associated with background are studied by varying the
fractions of signal (equivalent to the fractions of background),
i.e.~$w_{\rm sig}$ in Eq.~(\ref{likelihood3}). The fractions of signal
for the three sample groups are varied by one corresponding
statistical uncertainty. The largest differences from the nominal results
are assigned as the uncertainties.

The other source of potential bias arised from  the knowledge  of the background distributions.
We follow an alternative procedure by determining the background
shape with another two variables, $M^2_{K_{S}^{0}\pi^{+}}$ versus
$M^2_{\pi^{+}\pi^{0}}$, and change the smooth parameters in RooNDKeysPdf~\cite{Verkerke}.
This resulting change in results is small enough to be ignored and so we assign no uncertainty from this source.

\item[\lowercase\expandafter{\romannumeral4}]
Experimental effects.
To estimate the systematic uncertainty related to the difference in acceptance
between MC and data associated with the PID and tracking efficiencies, that is
$\gamma_{\epsilon}$ in Eq.~(\ref{MC-intergral-corrected}), the amplitude
fit is performed varying the PID and tracking efficiencies according to their
uncertainties.

\item[\lowercase\expandafter{\romannumeral5}]
Fit bias.
The amplitude analysis is performed on three-hundred data-sized signal MC samples and the
pulls, which are the normalized-residual distributions of the fit, are inspected to look for biases or
significant excursions from a normal distribution.
These studies indicate that the FFs of $D_{s}^{+} \to K^{*}(892)^{0}\pi^{+}$ and $D_{s}^{+} \to K^{*}(1410)^{0}\pi^{+}$ are slightly biased. In addition, the statistical uncertainties of the FF and the phase of $D_{s}^{+} \to K^{*}(892)^{0}\pi^{+}$ are underestimated. Therefore, we correct the biased FFs by the mean values of the pull distributions and scale the underestimated uncertainties by the widths of the pulls.
The systematic
uncertainty due to the correction is assigned as the uncertainty of the mean
value. An additional systematic uncertainty due to the normalization is taken
into account  by $\sqrt{2f\Delta f}$, where $f$ is the fitted width and
$\Delta f$ is its uncertainty~\cite{err_on_err1, err_on_err2}.
\end{itemize}

\begin{table*}[tp]
  \renewcommand\arraystretch{1.25}
  \centering
  \caption{Systematic uncertainties on the $\phi$ and FFs for each
    amplitude in units of the corresponding statistical uncertainties.  The sources are:
    (\lowercase\expandafter{\romannumeral1}) Fixed parameters in the amplitudes,
    (\lowercase\expandafter{\romannumeral2}) The R values,
    (\lowercase\expandafter{\romannumeral3}) Background,
    (\lowercase\expandafter{\romannumeral4}) Experimental effects,
    (\lowercase\expandafter{\romannumeral5}) Fit bias.
}
  \label{systematic-uncertainties}
  \begin{tabular}{lccccccc}
    \hline
    \multirow{2}{*}{Amplitude}&\multicolumn{7}{c}{Source}\cr
    & & \lowercase\expandafter{\romannumeral1} &\lowercase\expandafter{\romannumeral2} &\lowercase\expandafter{\romannumeral3} &\lowercase\expandafter{\romannumeral4} &\lowercase\expandafter{\romannumeral5}& Total   \\
    \hline
    $D_{s}^{+} \to K_{S}^{0}\rho^{+}$                        & FF     & 0.03 & 0.49 & 0.02 & 0.23 & 0.03 & 0.54  \\
    \hline
    \multirow{2}{*}{$D_{s}^{+} \to K_{S}^{0}\rho(1450)^{+}$} & $\phi$ & 0.34 & 0.38 & 0.15 & 0.16 & 0.06 & 0.56  \\
                                                             & FF     & 0.34 & 0.95 & 0.05 & 0.19 & 0.03 & 1.03  \\
    \hline
    \multirow{2}{*}{ $D_{s}^{+} \to K^{*}(892)^{0}\pi^{+}$}  & $\phi$ & 0.03 & 0.16 & 0.26 & 0.27 & 0.23 & 0.47  \\
                                                             & FF     & 0.06 & 0.29 & 0.10 & 0.21 & 0.18 & 0.42  \\
    \hline
    \multirow{2}{*}{$D_{s}^{+} \to K^{*}(892)^{+}\pi^{0}$ }  & $\phi$ & 0.03 & 0.33 & 0.70 & 0.21 & 0.06 & 0.80  \\
                                                             & FF     & 0.10 & 0.03 & 0.15 & 0.25 & 0.03 & 0.31  \\
    \hline
    \multirow{2}{*}{$D_{s}^{+} \to K^{*}(1410)^{0}\pi^{+}$}  & $\phi$ & 0.15 & 0.02 & 0.25 & 0.14 & 0.06 & 0.33  \\
                                                             & FF     & 0.14 & 0.22 & 0.13 & 0.09 & 0.03 & 0.31  \\
    \hline
  \end{tabular}
\end{table*}

\section{Branching fraction measurement of $D_{s}^{+} \to K^0_{S}\pi^{+}\pi^{0}$}
\label{BFSelection}
\begin{figure*}[htp]
\begin{center}
\includegraphics[width=0.90\textwidth]{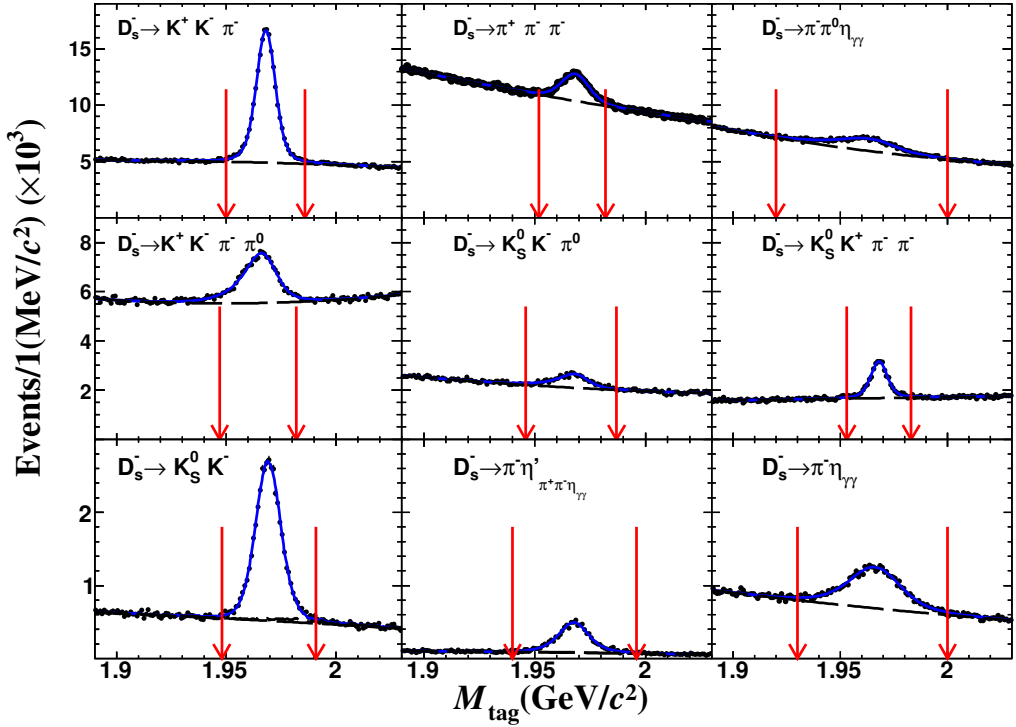}
\caption{Fits to the $M_{\rm tag}$ distributions of the ST candidates
         from the data sample at $\sqrt{s}=4.178$~GeV. The points with
         error bars are data, the blue solid lines are the total fits, and the black
         dashed lines are background. The pairs of red arrows denote the
         signal regions.
         }
\label{fit:Mass-data-Ds_4180}
\end{center}
\end{figure*}

With the selection criteria described in Sec.~\ref{ST-selection}, the best tag
candidate with $M_{\rm rec}$ closest to the $D_{s}^{\pm}$ nominal
mass~\cite{PDG} is chosen if there are multiple ST candidates. The yields for
various tag modes are listed in Table~\ref{ST-eff} and obtained by fitting the
corresponding $M_{\rm tag}$ distributions. As an example, the fits to the data
sample at $\sqrt s=4.178$~GeV are shown in Fig.~\ref{fit:Mass-data-Ds_4180}. In
the fits, the signal is modeled by an MC-simulated shape convolved with a
Gaussian function to take into account the data-MC resolution difference. The
background is described by a second-order Chebyshev function. MC studies show
that there is no significant peaking background in any tag mode, except for
$D^{-} \to K_{S}^{0} \pi^-$ and $D_{s}^{-} \to \eta\pi^+\pi^-\pi^-$ faking the
$D_{s}^{-} \to K_{S}^{0} K^-$ and $D_{s}^{-} \to \pi^-\eta^{\prime}$ tags,
respectively. Therefore, in the fits, the MC-simulated shapes of these two
peaking background sources are added to the background polynomial functions.
\begin{table*}[htbp]
  \caption{The ST yields for the samples collected at $\sqrt{s} =$ (I) 4.178, (II) 4.199-4.219,
    and (III) 4.23~GeV. The uncertainties are statistical.}\label{ST-eff}
    \begin{center}
      \begin{tabular}{lccc}
        \hline
        Tag mode                                     & (I) $N_{\rm ST}$                          & (II) $N_{\rm ST}$                       & (III) $N_{\rm ST}$    \\
        \hline
        $D_{s}^{-}\to K_{S}^{0}K^{-}$               & $\phantom{0}31668\pm315\phantom{0}$   & $18340\pm260\phantom{0}$            & $\phantom{0}6550\pm158\phantom{0}$  \\
        $D_{s}^{-}\to K^{+}K^{-}\pi^{-}$            & $135867\pm610\phantom{0}$             & $80417\pm507\phantom{0}$            & $28289\pm328\phantom{0}$ \\
        $D_{s}^{-}\to K_{S}^{0}K^{-}\pi^{0}$        & $\phantom{0}11284\pm512\phantom{0}$   & $\phantom{0}6729\pm462\phantom{0}$  & $\phantom{0}2144\pm218\phantom{0}$  \\
        $D_{s}^{-}\to K^{+}K^{-}\pi^{-}\pi^{0}$     & $\phantom{0}38421\pm767\phantom{0}$   & $22894\pm645\phantom{0}$            & $\phantom{0}7855\pm439\phantom{0}$  \\
        $D_{s}^{-}\to K_{S}^{0}K^{+}\pi^{-}\pi^{-}$ & $\phantom{0}15644\pm289\phantom{0}$   & $\phantom{0}8922\pm229\phantom{0}$  & $\phantom{0}3241\pm169\phantom{0}$  \\
        $D_{s}^{-}\to \pi^{-}\pi^{-}\pi^{+}$        & $\phantom{0}37702\pm853\phantom{0}$   & $21675\pm772\phantom{0}$            & $\phantom{0}7506\pm392\phantom{0}$  \\
        $D_{s}^{-}\to \pi^{-}\eta_{\gamma\gamma}$   & $\phantom{0}18070\pm560\phantom{0}$   & $10033\pm355\phantom{0}$            & $\phantom{0}3699\pm244\phantom{0}$  \\
        $D_{s}^{-}\to \pi^{-}\pi^{0}\eta_{\gamma\gamma}$
                                                    & $\phantom{0}40862\pm1313$             & $25877\pm1823$                      & $10659\pm1060$\\
        $D_{s}^{-}\to \pi^{-}\eta_{\pi^{+}\pi^{-}\eta_{\gamma\gamma}}^{'}$
                                                    & $\phantom{00}7773\pm143\phantom{0}$   & $\phantom{0}4464\pm111\phantom{0}$  & $\phantom{0}1676\pm74\phantom{00}$   \\
        \hline
      \end{tabular}
    \end{center}
\end{table*}

Once a tag mode is identified, we search for the signal decay
$D_{s}^{+} \to K_S^{0}\pi^{+}\pi^{0}$. In the case of multiple candidates, the
DT candidate with the average mass, $(M_{\rm sig}+M_{\rm tag})/2$,
closest to the $D_{s}^{\pm}$ nominal mass is retained.

To measure the BF, we start from the following equations for a single tag mode:
\begin{eqnarray}\begin{aligned}
  N_{\text{tag}}^{\text{ST}} = 2N_{D_{s}^{+}D_{s}^{-}}\mathcal{B}_{\text{tag}}\epsilon_{\text{tag}}^{\text{ST}}\,, \label{eq-ST}
\end{aligned}\end{eqnarray}
\begin{equation}
    N_{\text{tag,sig}}^{\text{DT}}=2N_{D_{s}^{+}D_{s}^{-}}\mathcal{B}_{\text{tag}}\mathcal{B}_{\text{sig}}\epsilon_{\text{tag,sig}}^{\text{DT}}\,,
  \label{eq-DT}
\end{equation}
where $N_{D_{s}^{+}D_{s}^{-}}$ is the total number of $D_{s}^{*\pm}D_{s}^{\mp}$
pairs produced from the $e^{+}e^{-}$ collisions; $N_{\text{tag}}^{\text{ST}}$ is
the ST yield for the tag mode; $N_{\text{tag,sig}}^{\text{DT}}$ is the DT yield;
$\mathcal{B}_{\text{tag}}$ and $\mathcal{B}_{\text{sig}}$ are the BFs of the
tag and signal modes, respectively; $\epsilon_{\text{tag}}^{\text{ST}}$ is the
ST efficiency to reconstruct the tag mode; and $\epsilon_{\text{tag,sig}}^{\text{DT}}$
is the DT efficiency to reconstruct both the tag and the signal decay modes. In
the case of more than one tag modes and sample groups,
\begin{eqnarray}
\begin{aligned}
  \begin{array}{lr}
    N_{\text{total}}^{\text{DT}}=\Sigma_{\alpha, i}N_{\alpha,\text{sig},i}^{\text{DT}}   = \mathcal{B}_{\text{sig}}
 \Sigma_{\alpha, i}2N_{D_{s}^{+}D_{s}^{-}}\mathcal{B}_{\alpha}\epsilon_{\alpha,\text{sig}, i}^{\text{DT}}\,,
  \end{array}
  \label{eq-DTtotal}
\end{aligned}
\end{eqnarray}
where $\alpha$ represents tag modes in the $i$th sample group.
By isolating $\mathcal{B}_{\text{sig}}$, we find
\begin{eqnarray}\begin{aligned}
  \mathcal{B}_{\text{sig}} =
  \frac{N_{\text{total}}^{\text{DT}}}{ \mathcal{B}_{K_S^0\to\pi^+\pi^-}\mathcal{B}_{\pi^0\to\gamma\gamma}\begin{matrix}\sum_{\alpha, i} N_{\alpha, i}^{\text{ST}}\epsilon^{\text{DT}}_{\alpha,\text{sig},i}/\epsilon_{\alpha,i}^{\text{ST}}\end{matrix}}\,.
\end{aligned}\end{eqnarray}
where $N_{\alpha,i}^{\text{ST}}$ and $\epsilon_{\alpha,i}^{\text{ST}}$ are
obtained from the data and generic MC samples, respectively, while
$\epsilon_{\alpha,\text{sig},i}^{\text{DT}}$ is determined with signal MC
samples, where $D_{s}^{+} \to K_S^{0}\pi^{+}\pi^{0}$ events are generated
according to the results of the  amplitude analysis.  The two branching ratios
$ \mathcal{B}_{K_S^0\to\pi^+\pi^-}$ and $\mathcal{B}_{\pi^0\to\gamma\gamma}$
have been introduced to account for the fact that the signal is reconstructed through
these decays.

\begin{figure}[!htbp]
  \centering
  \includegraphics[width=0.6\textwidth]{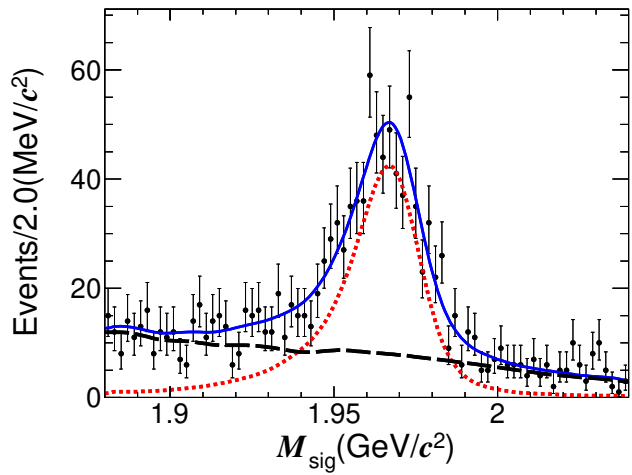}
 \caption{Fit to the $M_{\rm sig}$ distribution of the DT candidates from the
   data samples  at $\sqrt{s}= 4.178$-$4.226$~GeV. The data are represented by
   points with error bars, the total fit by the blue solid line, and the fitted
   signal and the fitted background by the red dotted and the black dashed
   lines, respectively.
 }
  \label{DT-fit}
\end{figure}

The DT yield $N_{\text{total}}^{\text{DT}}$ is found to be $666\pm37$ from the
fit to the $M_{\rm sig}$ distribution of the selected $D^+_s\to K^0_S\pi^+\pi^0$
candidates. The fit result is shown in Fig.~\ref{DT-fit}. In the fit, the signal
shape is described by an MC-simulated shape convolved with a Gaussian function
to take into account the data-MC resolution difference. The background shape is
described by an MC-simulated shape, which includes the small peaking
background~($2.1\%$) that is mainly from
$D_{s}^{+} \to \pi^{+}\pi^{+}\pi^{-}\pi^{0}$ decays. The width of the Gaussian
function is fixed to be $1.9\pm 1.1$~MeV/$c^{2}$, which is extracted from the
control sample of  $D_{s}^{+} \to K_{S}^{0}K^{+}\pi^{0}$ decays. Note that
the DT yield is larger than the fit yields of Fig.~\ref{fig:fit_Ds} since
the selection for the BF measurement is looser than that for the amplitude analysis
and no kinematic fit is applied in the BF measurement.

We take the differences in pion tracking efficiency between data and MC
simulation into account, and apply a correction to the MC signal efficiency of $+0.3\%$.
The differences in PID efficiency are negligible.
The BF is determined to be
$\mathcal{B}(D_{s}^{+} \to K_{S}^0\pi^{+}\pi^{0})=(5.43\pm0.30_{\rm stat}\pm 0.15_{\rm syst})\times 10^{-3}$.

In order to test $CP$ conservation in the decay,
the BFs are measured separately for the charge-conjugated modes.
The BFs of
$D_{s}^{+} \to K_{S}^0\pi^{+}\pi^{0}$ and $D_{s}^{-} \to K_{S}^0\pi^{-}\pi^{0}$
are measured to be $(5.33\pm0.41_{\rm stat}\pm 0.15_{\rm syst})\times 10^{-3}$
and $(5.63\pm0.44_{\rm stat}\pm 0.16_{\rm syst})\times 10^{-3}$, respectively.
From these measurements the asymmetry
of the BFs is determined to be $(2.7\pm5.5_{\rm stat}\pm0.9_{\rm syst})\%$ by
$A_{CP}=\frac{B(D_{s}^{+})-B(D_{s}^{-})}{B(D_{s}^{+})+B(D_{s}^{-})}$, where
$B(D_{s}^{+(-)})$ is the BF of the decay $D_{s}^{+(-)} \to K_{S}^0\pi^{+(-)}\pi^{0}$.
Hence, no $CP$ violation is observed.
Note that the systematic uncertainties related to $K_{S}^0$ and $\pi^{0}$
reconstructions cancel in the $A_{CP}$ calculation.

The following sources of the systematic uncertainties are taken into account
for the BF measurement.
\begin{itemize}
\item Signal shape. The systematic uncertainty due to the signal shape is
  studied by repeating the fit with an alternative width of the convolved
  Gaussian. This width is varied according to the uncertainty of the control
  sample.

  \item Background shape. Since $q\bar{q}$ or non-$D_{s}^{*\pm}D_{s}^{\mp}$
  open charm are the major background sources, we alter the MC shapes by varying
  the relative fractions of the background from $q\bar{q}$ or
  non-$D_{s}^{*+}D_{s}^{-}$ open charm by $\pm 30\%$. The largest change is
  taken as the corresponding systematic uncertainty.

\item $\pi^{+}$ tracking/PID efficiency. The $\pi^{+}$ tracking and PID
  efficiencies are studied with $e^+e^-\to K^+K^-\pi^+\pi^-$ events. The
  data-MC efficiency ratios of the $\pi^{+}$ tracking and PID efficiencies are
  $1.003\pm0.002$ and $1.000\pm0.002$, respectively. After multiplying the
  signal efficiencies by the factor $1.003$, we assign $0.2\%$ and $0.2\%$ as
  the systematic uncertainties arising from $\pi^{+}$ PID and tracking,
  respectively.

\item $K_S^0$ reconstruction. The systematic uncertainty from the $K^0_S$ reconstruction
  efficiency is assigned to be $1.5\%$, determined  from studying a control
  sample of $\psi(3770)\to D\bar{D}$ events containing hadronic $D$ decays.

\item $\pi^{0}$ reconstruction. A control sample of the process
  $e^+e^-\to K^+K^-\pi^+\pi^-\pi^0$ is used to study the uncertainty
  due to $\pi^0$ reconstruction, which is assigned as 2.0\%.

\item MC statistics. The uncertainty due to the limited MC statistics is
  obtained by
  $\sqrt{\begin{matrix} \sum_{i} (f_{i}\frac{\delta_{\epsilon_{i}}}{\epsilon_{i}}\end{matrix}})^2$,
  where $f_{i}$ is the tag yield fraction, and $\epsilon_{i}$ and
  $\delta_{\epsilon_{i}}$ are the signal efficiency and the corresponding
  uncertainty of tag mode $i$, respectively.

\item Dalitz model. The uncertainty from the Dalitz model is estimated by
  varying the Dalitz model parameters based on their error matrix. The
  distribution of 600 efficiencies resulting from this variation  is fitted by a Gaussian function
  and the deviation from the nominal mean value is taken as an uncertainty.

\item Peaking background. The uncertainty of the peaking background is about
8\% and corresponds to only one event in the $D_s^+\to K_{S}^{0}\pi^+\pi^0$
decay. Therefore the associated uncertainty in the BF measurement is negligible.
\end{itemize}

All of the systematic uncertainties are summarized in Table~\ref{BF-Sys}.
Adding them in quadrature gives a total systematic uncertainty in the BF
measurement of 2.8\%.
\begin{table}[htbp]
  \caption{Systematic uncertainties in the BF measurement.}
  \label{BF-Sys}
  \begin{center}
    \begin{tabular}{lccc}
      \hline
      Source   & Sys. Uncertainty (\%)\\
      \hline
      Signal shape                        & 0.8 \\
      Background shape                    & 0.5 \\
      $\pi^{+}$ PID efficiency            & 0.2 \\
      $\pi^{+}$ tracking efficiency       & 0.2 \\
      $K_S^0$ reconstruction              & 1.5 \\
      $\pi^0$ reconstruction              & 2.0 \\
      MC statistics                       & 0.3 \\
      Dalitz model                        & 0.8 \\
      \hline
      Total                               & 2.8 \\
      \hline
    \end{tabular}
  \end{center}
\end{table}

\section{Summary}
An amplitude analysis has been performed for the decay
$D_{s}^{+} \to K_{S}^0\pi^{+}\pi^{0}$. The results for the  FFs and phases among the
different intermediate processes are listed in Table~\ref{fit-result}.
After calculating a detection efficiency that accounts for the variation of decays over phase space
found in the amplitude analysis, the
BF for the decay $D^+_s\to K^0_S\pi^+\pi^0$ is measured to be
$(5.43\,\pm\,0.30_{\rm stat}\,\pm\,0.15_{\rm syst})\times 10^{-3}$ with an
improved precision by about a factor of 3 compared to the PDG value~\cite{PDG}.
The BFs for the intermediate processes are calculated with
$\mathcal{B}_{i} = {\rm FF}_{i} \times \mathcal{B}(D_{s}^{+} \to K_{S}^0\pi^{+}\pi^{0})$
and listed in Table~\ref{inter-processes}. Assuming
$\mathcal{B}(K^0\to K^0_S)=0.5$, we determine
$\mathcal{B}(D_{s}^{+} \to K^{0}\rho^{+}) = (5.46\,\pm\,0.84_{\rm stat}\,\pm\,0.44_{\rm syst})\times 10^{-3}$,
$\mathcal{B}(D_{s}^{+} \to K^{*}(892)^{0}\pi^{+}) = (2.71\,\pm\,0.72_{\rm stat}\,\pm\,0.30_{\rm syst})\times 10^{-3}$, and
$\mathcal{B}(D_{s}^{+} \to K^{*}(892)^{+}\pi^{0}) = (0.75\,\pm\,0.24_{\rm stat}\,\pm\,0.06_{\rm syst})\times 10^{-3}$.
Our results are valuable for a deeper understanding of quark flavor SU(3) symmetry, SU(3)
breaking effects, and other related theoretical issues.

These results can be compared to the current theoretical predictions~\cite{YLWu, HYCheng, PRD84-074019}. The predictions in Ref~\cite{YLWu} are consistent with our results, but their large uncertainties make the comparisons less conclusive. The calculations in Ref~\cite{HYCheng} have small uncertainties, while the predicted  $\mathcal{B}(D_{s}^{+} \to K^{0}\rho^{+})$ is over five standard deviations off the measured one. The predictions in Ref~\cite{PRD84-074019} have moderate uncertainties and match our measurements in principle, but the predicted $\mathcal{B}(D_{s}^{+} \to K^{*}(892)^{+}\pi^{0})$ is only marginally consistent with our measurement. Based on the current experimental and theoretical precisions, it is difficult to draw a definite conclusion to discriminate between models yet.

The asymmetry for the BFs of the decays $D_{s}^{+} \to K_{S}^0\pi^{+}\pi^{0}$
and $D_{s}^{-} \to K_{S}^0\pi^{-}\pi^{0}$ is determined to be
$(2.7\pm5.5_{\rm stat}\pm0.9_{\rm syst})\%$. No evidence for $CP$ violation is
found.

\begin{table}[htbp]
  \caption{The BFs for various intermediate processes with the final state
    $K_{S}^0\pi^{+}\pi^{0}$. The first and second uncertainties are
    statistical and systematic, respectively.}\label{inter-processes}
  \begin{center}
    \begin{tabular}{lc}
      \hline
      Intermediate process & BF ($10^{-3}$)\\
      \hline
      $D_{s}^{+} \rightarrow K_{S}^{0}\rho^{+}$        & $ 2.73 \pm 0.42 \pm 0.22$  \\
      $D_{s}^{+} \rightarrow K_{S}^{0}\rho(1450)^{+}$  & $ 1.11 \pm 0.24 \pm 0.24$  \\
      $D_{s}^{+} \rightarrow K^{*}(892)^{0}\pi^{+}$    & $ 0.45 \pm 0.12 \pm 0.05$  \\
      $D_{s}^{+} \rightarrow K^{*}(892)^{+}\pi^{0}$    & $ 0.25 \pm 0.08 \pm 0.02$  \\
      $D_{s}^{+} \rightarrow K^{*}(1410)^{0}\pi^{+}$   & $ 0.18 \pm 0.09 \pm 0.03$  \\
      \hline
    \end{tabular}
  \end{center}
\end{table}

\acknowledgments
The BESIII collaboration thanks the staff of BEPCII and the IHEP computing center for their strong support. This work is supported in part by National Key Research and Development Program of China under Contracts Nos. 2020YFA0406300, 2020YFA0406400; National Natural Science Foundation of China (NSFC) under Contracts Nos. 11625523, 11635010, 11735014, 11775027, 11822506, 11835012, 11875054, 11935015, 11935016, 11935018, 11961141012; the Chinese Academy of Sciences (CAS) Large-Scale Scientific Facility Program; Joint Large-Scale Scientific Facility Funds of the NSFC and CAS under Contracts Nos. U1732263, U1832207, U2032104; CAS Key Research Program of Frontier Sciences under Contracts Nos. QYZDJ-SSW-SLH003, QYZDJ-SSW-SLH040; 100 Talents Program of CAS; INPAC and Shanghai Key Laboratory for Particle Physics and Cosmology; ERC under Contract No. 758462; European Union Horizon 2020 research and innovation programme under Contract No. Marie Sklodowska-Curie grant agreement No 894790; German Research Foundation DFG under Contracts Nos. 443159800, Collaborative Research Center CRC 1044, FOR 2359, FOR 2359, GRK 214; Istituto Nazionale di Fisica Nucleare, Italy; Ministry of Development of Turkey under Contract No. DPT2006K-120470; National Science and Technology fund; Olle Engkvist Foundation under Contract No. 200-0605; STFC (United Kingdom); The Knut and Alice Wallenberg Foundation (Sweden) under Contract No. 2016.0157; The Royal Society, UK under Contracts Nos. DH140054, DH160214; The Swedish Research Council; U. S. Department of Energy under Contracts Nos. DE-FG02-05ER41374, DE-SC-0012069.

\bibliographystyle{JHEP}
\bibliography{references}

\providecommand{\href}[2]{#2}\begingroup\raggedright\begin{thebibliography}{10}

\bibitem{PDG}
{\scshape Particle Data Group} collaboration, \emph{{Review of Particle
  Physics}}, \href{https://doi.org/10.1093/ptep/ptaa104}{\emph{PTEP} {\bfseries
  2020} (2020) 083C01}.

\bibitem{PRD79-034016}
B.~Bhattacharya and J.~L. Rosner, \emph{{Decays of charmed mesons to $PV$ final
  states}}, \href{https://doi.org/10.1103/PhysRevD.79.034016}{\emph{Phys. Rev.
  D} {\bfseries 79} (2009) 034016}.

\bibitem{PRD81-074021}
H.-Y. Cheng and C.-W. Chiang, \emph{Two-body hadronic charmed meson decays},
  \href{https://doi.org/10.1103/PhysRevD.81.074021}{\emph{Phys. Rev. D}
  {\bfseries 81} (2010) 074021}.

\bibitem{PRD84-074019}
F.-S. Yu, X.-X. Wang and C.-D. L\"u, \emph{Nonleptonic two-body decays of
  charmed mesons},
  \href{https://doi.org/10.1103/PhysRevD.84.074019}{\emph{Phys. Rev. D}
  {\bfseries 84} (2011) 074019}.

\bibitem{YLWu}
{Wu, Yue-Liang}, {Zhong, Ming} and {Zhou, Yu-Feng}, \emph{{Exploring final
  state hadron structure and SU(3) flavor symmetry breaking effects in $D\to
  PP$ and $D\to PV$ decays}},
  \href{https://doi.org/10.1140/epjc/s2005-02302-2}{\emph{Eur. Phys. J. C}
  {\bfseries 42} (2005) 391}.

\bibitem{HYCheng}
H.-Y. Cheng and C.-W. Chiang, \emph{{Revisiting $CP$ violation in
  $D\ensuremath{\rightarrow}PP$ and $VP$ decays}},
  \href{https://doi.org/10.1103/PhysRevD.100.093002}{\emph{Phys. Rev. D}
  {\bfseries 100} (2019) 093002}.

\bibitem{CLEO-BF}
{\scshape CLEO} collaboration, \emph{Measurement of the pseudoscalar decay
  constant ${f}_{{D}_{s}}$ using
  ${D}_{s}^{+}\ensuremath{\rightarrow}{\ensuremath{\tau}}^{+}\ensuremath{\nu}$,
  ${\ensuremath{\tau}}^{+}\ensuremath{\rightarrow}{\ensuremath{\rho}}^{+}\overline{\ensuremath{\nu}}$
  decays}, \href{https://doi.org/10.1103/PhysRevD.80.112004}{\emph{Phys. Rev.
  D} {\bfseries 80} (2009) 112004}.

\bibitem{Ablikim:2009aa}
M.~Ablikim, Z.~An, J.~Bai, N.~Berger, J.~Bian, X.~Cai et~al., \emph{Design and
  construction of the {BESIII} detector},
  \href{https://doi.org/https://doi.org/10.1016/j.nima.2009.12.050}{\emph{Nucl.
  Instrum. Meth. A} {\bfseries 614} (2010) 345}.

\bibitem{Ablikim:2019hff}
{\scshape BESIII} collaboration, \emph{{Future physics programme of BESIII}},
  \href{https://doi.org/10.1088/1674-1137/44/4/040001}{\emph{Chin. Phys. C}
  {\bfseries 44} (2020) 040001}
  [\href{https://arxiv.org/abs/1912.05983}{{\ttfamily arXiv:1912.05983}}].

\bibitem{Yu:IPAC2016-TUYA01}
C.~Yu et~al., \emph{{BEPCII} {P}erformance and {B}eam {D}ynamics {S}tudies on
  {L}uminosity},  in \emph{Proc. of International Particle Accelerator
  Conference (IPAC'16), Busan, Korea, May 8-13, 2016}, no.~7 in International
  Particle Accelerator Conference, (Geneva, Switzerland), pp.~1014--1018,
  JACoW, June, 2016,
  \href{https://doi.org/doi:10.18429/JACoW-IPAC2016-TUYA01}{DOI}.

\bibitem{DsStrDs}
{\scshape CLEO} collaboration, \emph{{Measurement of charm production cross
  sections in ${e}^{+}{e}^{\ensuremath{-}}$ annihilation at energies between
  3.97 and 4.26~GeV}},
  \href{https://doi.org/10.1103/PhysRevD.80.072001}{\emph{Phys. Rev. D}
  {\bfseries 80} (2009) 072001}.

\bibitem{GEANT4}
{\scshape GEANT4} collaboration, \emph{{GEANT4 -- a simulation toolkit}},
  \href{https://doi.org/10.1016/S0168-9002(03)01368-8}{\emph{Nucl. Instrum.
  Meth. A} {\bfseries 506} (2003) 250}.

\bibitem{KKMC}
S.~Jadach, B.~F.~L. Ward and Z.~Wa\ifmmode~\mbox{\c{}}\else \c{}\fi{}s,
  \emph{Coherent exclusive exponentiation for precision monte carlo
  calculations}, \href{https://doi.org/10.1103/PhysRevD.63.113009}{\emph{Phys.
  Rev. D} {\bfseries 63} (2001) 113009}.

\bibitem{EVTGEN1}
D.~Lange, \emph{{The EvtGen particle decay simulation package}},
  \href{https://doi.org/10.1016/S0168-9002(01)00089-4}{\emph{Nucl. Instrum.
  Meth. A} {\bfseries 462} (2001) 152}.

\bibitem{EVTGEN2}
R.-G. Ping, \emph{Event generators at {BESIII}},
  \href{https://doi.org/10.1088/1674-1137/32/8/001}{\emph{Chin. Phys. C}
  {\bfseries 32} (2008) 599}.

\bibitem{LUNDCHARM1}
J.~C. Chen, G.~S. Huang, X.~R. Qi, D.~H. Zhang and Y.~S. Zhu, \emph{{Event
  generator for $J/\ensuremath{\psi}$ and $\ensuremath{\psi}(2S)$ decay}},
  \href{https://doi.org/10.1103/PhysRevD.62.034003}{\emph{Phys. Rev. D}
  {\bfseries 62} (2000) 034003}.

\bibitem{LUNDCHARM2}
R.-L. Yang, R.-G. Ping and H.~Chen, \emph{{Tuning and validation of the
  Lundcharm model with $J/\psi$ decays}},
  \href{https://doi.org/10.1088/0256-307X/31/6/061301}{\emph{Chin. Phys. Lett.}
  {\bfseries 31} (2014) 061301}.

\bibitem{PHOTOS}
E.~Richter-Was, \emph{{QED bremsstrahlung in semileptonic B and leptonic $\tau$
  decays}},
  \href{https://doi.org/https://doi.org/10.1016/0370-2693(93)90062-M}{\emph{Phys.
  Lett. B} {\bfseries 303} (1993) 163 }.

\bibitem{Verkerke}
W.~Verkerke and D.~P. Kirkby, \emph{RooFit Users Manual v2.07}. 2006,
  \href{https://doi.org/http://roofit.sourceforge.net}{http://roofit.sourceforge.net}.

\bibitem{PhysRevLett.21.244}
G.~J. Gounaris and J.~J. Sakurai, \emph{Finite-width corrections to the
  vector-meson-dominance prediction for
  $\ensuremath{\rho}\ensuremath{\rightarrow}{e}^{+}{e}^{\ensuremath{-}}$},
  \href{https://doi.org/10.1103/PhysRevLett.21.244}{\emph{Phys. Rev. Lett.}
  {\bfseries 21} (1968) 244}.

\bibitem{covariant-tensors}
{B.S. Zou} and {D.V. Bugg}, \emph{Covariant tensor formalism for partial-wave
  analyses of $\psi$ decay to mesons},
  \href{https://doi.org/10.1140/epja/i2002-10135-4}{\emph{Eur. Phys. J. A}
  {\bfseries 16} (2003) 537}.

\bibitem{err_on_err1}
R.~Barlow, \emph{{Systematic Errors: facts and fictions}}, {\emph{arXiv}
  {\bfseries hep-ex} (2002) 0207026}
  [\href{https://arxiv.org/abs/hep-ex/0207026}{{\ttfamily hep-ex/0207026}}].

\bibitem{err_on_err2}
R.~D. Cousins and V.~L. Highland, \emph{Incorporating systematic uncertainties
  into an upper limit},
  \href{https://doi.org/https://doi.org/10.1016/0168-9002(92)90794-5}{\emph{Nucl.
  Instrum. Meth. A} {\bfseries 320} (1992) 331}.

\end{thebibliography}\endgroup

\end{document}